\documentclass[prl,aps,twocolumn,superscriptaddress]{revtex4-2}
\usepackage{graphicx}
\usepackage{amsmath}
\usepackage{amssymb}
\usepackage{xcolor}
\usepackage{slashed}
\usepackage{hyperref}
\usepackage{cleveref}
\hypersetup{pdftitle={},pdfcreator={},linkcolor=[rgb]{0.15,0.35,0.75},colorlinks=true,citecolor=[rgb]{0.675,0,0.2},urlcolor=[rgb]{0.15,0.35,0.65}}
\allowdisplaybreaks

\begin{document}

	\title{Recoil Geometry Unmasks Gluon Saturation in Forward $Z^0$ Production}
	
	\author{Wanchen Li}
	\email{wanchenli@fudan.edu.cn}
	\affiliation{Department of Physics, Center for Field Theory and Particle Physics, Key Laboratory of Nuclear Physics and Ion-beam Application (MOE), Fudan University, Shanghai 200433, China}
	
	\author{Ding Yu Shao}
	\email{dyshao@fudan.edu.cn}
	\affiliation{Department of Physics, Center for Field Theory and Particle Physics, Key Laboratory of Nuclear Physics and Ion-beam Application (MOE), Fudan University, Shanghai 200433, China}
	\affiliation{Shanghai Research Center for Theoretical Nuclear Physics,  Fudan University, Shanghai 200438, China}
	\affiliation{Center for High Energy Physics, Peking University, Beijing 100871, China}
	\affiliation{Southern Center for Nuclear-Science Theory (SCNT), Institute of Modern Physics, Chinese Academy of Sciences, Huizhou, Guangdong 516000, China}
	
	\author{Shu-Yi Wei}
	\email{shuyi@sdu.edu.cn}
	\affiliation{Key Laboratory of Particle Physics and Particle Irradiation (MOE), Institute of Frontier and Interdisciplinary Science, Shandong University (Qingdao), Shandong 266237, China}
	
	\author{Jian Zhou}
	\email{jzhou@sdu.edu.cn}
	\affiliation{Key Laboratory of Particle Physics and Particle Irradiation (MOE), Institute of Frontier and Interdisciplinary Science, Shandong University (Qingdao), Shandong 266237, China}
	\affiliation{Southern Center for Nuclear-Science Theory (SCNT), Institute of Modern Physics, Chinese Academy of Sciences, Huizhou, Guangdong 516000, China}

	\begin{abstract}
		Gluon saturation produces characteristic transverse-momentum broadening in nuclei, but QCD radiation largely washes out this signature.
		We show that fiducial recoil subtraction turns detector acceptance into a transverse-momentum projector that unmasks the broadening in forward $Z^0$ production.  
		Subtracting the hadronic recoil measured in a chosen rapidity interval from the boson transverse momentum defines a residual momentum.
		At leading power, the radiative recoil in this interval cancels, while the residual momentum retains sensitivity to the small-$x$ nuclear field.
		Combining a CGC description of the small-$x$ target with soft-collinear effective theory (SCET) resummation for finite rapidity coverage, we find that a benchmark rapidity coverage $|\eta^{\rm lab}|<2.5$ lowers the effective hard scale from $M_Z\simeq 91.2$ GeV to about $7.5~\mathrm{GeV}$ of the Sudakov evolution.
		Increasing the saturation scale broadens the residual-momentum distribution and weakens recoil alignment, whereas wider coverage makes the proton--nucleus separation clearer in both observables. 
		Detector geometry thus provides tunable control over perturbative recoil, enabling a probe of nonlinear small-$x$ QCD.
	\end{abstract}

	\maketitle
	
	\emph{\textbf{Introduction.}}
	A central question in high-energy QCD is how the growth of small-$x$ gluon densities is tamed and how that nonlinear dynamics is recorded in transverse momentum flow.
	The Color Glass Condensate (CGC) organizes this physics around the saturation scale $Q_s(x)$, which sets the characteristic transverse momentum generated by multiple scattering in the dense small-$x$ field~\cite{Gribov:1984tu, Mueller:1985wy, McLerran:1993ni, McLerran:1993ka, Mueller:1999wm, Gelis:2010nm}.
	A nucleus enhances this scale and imprints nonlinear dynamics on the recoil: relative to a dilute proton, saturation broadens the recoil transverse-momentum distribution and weakens away-side alignment~\cite{Schafer:2013mza, Albacete:2018ruq}.
	Forward dihadron measurements at RHIC and dijet measurements at the LHC map this nuclear modification~\cite{Adare:2011sc, Aaboud:2019oop}, while CGC calculations connect it to the recoil geometry of the dense target~\cite{Albacete:2010pg, Lappi:2012nh, Akcakaya:2012si, Stasto:2012ru, Kutak:2012rf, Kotko:2015ura, vanHameren:2016ftb,Stasto:2018rci, vanHameren:2019ysa, Al-Mashad:2022zbq, vanHameren:2023oiq, Caucal:2025zkl}.
	The central experimental task is to resolve the nuclear contribution within the total measured recoil and thereby identify the transverse momentum transferred by the small-$x$ field.
	
	Conventional transverse-momentum imbalance observables $q_T$ combine the nuclear momentum transfer and perturbative radiation in a single distribution~\cite{Stasto:2018rci, vanHameren:2019ysa, BermudezMartinez:2019anj, Blanco:2019qbm, Benic:2022ixp, Marquet:2025jdr, Gao:2026azd}.
	In the small-recoil region \(q_T\ll Q\), this mixing has a sharp parametric origin: soft and collinear radiation generates Sudakov logarithms of the form \(\alpha_s^n\ln^{2n}(Q^2/q_T^2)\) and opens radiative phase space between $q_T$ and $Q$~\cite{Mueller:2012uf, Mueller:2013wwa}.
	Joint resummation of small-\(x\) and Sudakov logarithms determines the evolution of the combined distribution~\cite{Watanabe:2015yca, Xiao:2017yya, Zhou:2018lfq, vanHameren:2020rqt, Balitsky:2020jzt, Hentschinski:2021lsh, Goda:2022wsc, Mukherjee:2023snp, Altinoluk:2024vgg, Duan:2024nlr, vanHameren:2025hyo, Balitsky:2026lop}.
	At NLO, the interplay between small-$x$ dynamics and Sudakov radiation has been studied for back-to-back dijets in DIS and photoproduction and for dihadron production in DIS~\cite{Caucal:2022ulg,  Taels:2022tza, Caucal:2023nci, Caucal:2023fsf, Caucal:2024nsb}.
	Forward \(Z^0\) production realizes a wide hierarchy, \(Q\simeq M_Z\gg Q_s\), and makes the radiative competition especially pronounced.
	Separating the momentum sources requires one additional piece of information: where the recoil flows relative to a declared fiducial region.  
	The inclusive \(Z^0\) transverse momentum is balanced by recoil over all rapidities and therefore carries no information about where that recoil flows.
	
	\begin{figure}[t]
		\centering
		\includegraphics[width=\columnwidth]{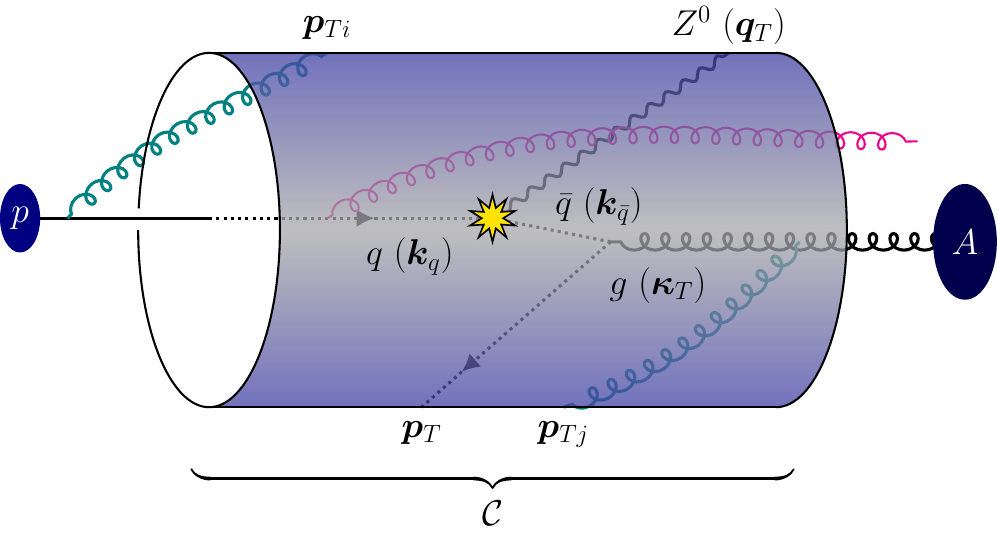}
		\caption{\label{fig:kinematics}
			Recoil geometry in forward \(p+A\to Z^0+X\).
			The projectile quark carrying transverse momentum \(\boldsymbol k_q\) annihilates with a small-\(x\) antiquark carrying \(\boldsymbol k_{\bar q}\), generated from the nuclear gluon field with a companion of transverse momentum \(\boldsymbol p_T\).  
			The shaded cylinder denotes the fiducial coverage \(\mathcal C\): recorded hadronic transverse momenta form the fiducial vector recoil, while flow beyond the coverage remains in the residual momentum.}
	\end{figure}
	
	In this Letter, we exploit the rapidity distribution of the recoil to construct an acceptance-dependent observable for forward \(p+A\to Z^0+X\), taking \(p+p\) as the baseline.
	For a declared fiducial coverage \(\mathcal C\), excluding the reconstructed boson, we form the fiducial vector recoil \(\boldsymbol l_T\) from all recorded hadrons inside \(\mathcal C\). 
	This observable can be viewed as a finite-coverage vector angularity~\cite{Bijl:2023dux} with a step-function rapidity weight.
	We denote its magnitude by $l_T\equiv|\boldsymbol l_T|$ and define the residual transverse-momentum imbalance as $\boldsymbol k_T\equiv\boldsymbol q_T+\boldsymbol l_T$.
	At leading power in \(q_T/Q\) and for a fixed central coverage, soft radiation recorded inside \(\mathcal C\) shifts \(\boldsymbol q_T\) and \(\boldsymbol l_T\) oppositely and therefore cancels from \(\boldsymbol k_T\), whereas radiation outside \(\mathcal C\) remains.
	At this accuracy, the residual distribution retains the small-\(x\) nuclear contribution and projectile transverse momentum, while only radiation outside \(\mathcal C\) contributes to its hard-scale radiative broadening.
	Detector coverage thus acts as a tunable projector that reduces the radiative background and exposes target-dependent broadening.
	Forward \(Z^0\) production provides a particularly clean setting: the color-neutral boson fixes the hard scale without a fragmentation function, while forward kinematics probes the small-\(x\) target sea-quark distribution~\cite{Hautmann:2012sh, Marquet:2019ltn}.
	
	Joint event-shape and \(q_T\) resummation provide the starting point~\cite{Stewart:2010tn, Kang:2012zr, Kang:2013wca, Procura:2014cba, Lustermans:2019plv, Monni:2019yyr, Makris:2020ltr, Boglione:2020auc, Boglione:2021wov, Boglione:2023duo, Fang:2025dee}.
	Incorporating finite rapidity cuts~\cite{Michel:2018hui}, we derive the joint recoil distribution in a CGC-matched soft-collinear effective theory framework~\cite{Bauer:2000ew, Bauer:2000yr, Bauer:2001ct, Bauer:2001yt}.
	The CGC beam function encodes the target transverse momentum, while the fiducial measurement partitions the Drell--Yan Sudakov factor between recoil recorded inside \(\mathcal C\) and recoil that flows outside it.
	For the laboratory-frame coverage $-2.5<\eta^{\rm lab}<2.5$, the uncovered double-logarithmic phase space is characterized by an effective residual Sudakov scale $Q_{\rm eff}\simeq7.5~\mathrm{GeV}$, far below the $Z^0$ mass.
	Increasing \(Q_s\) strengthens the nuclear kick, broadening the residual-momentum distribution and weakening recoil alignment, whereas widening the coverage at fixed \(x_A\) lowers \(Q_{\rm eff}\) and makes these proton--nucleus differences more visible.
	Target and acceptance scans therefore probe the nonlinear and radiative scales in complementary ways; the correlated response of both observables to these scans, rather than a single broadened spectrum, provides a clear test of saturation.
	
	\emph{\textbf{Forward $Z^0$ production in the CGC.}}
	In forward \(p+A\) collisions, the dilute--dense hybrid framework describes the proton as a large-\(x\) projectile and the nucleus as a dense small-\(x\) target~\cite{Gelis:2002nn, Altinoluk:2011qy,Chirilli:2011km}.
	The scattering is dominated by a large-$x$ quark from the proton annihilating with a small-$x$ sea antiquark from the nucleus, alongside the corresponding crossed channel. 
	This annihilation is localized at the short-distance scale $1/Q$ and involves no colored hard final-state leg. 
	The nuclear transverse-momentum dependence is therefore encoded in the target distributions, while the Born hard coefficient function remains identical to the tree-level Drell--Yan result~\cite{Kopeliovich:2000fb, Gelis:2002fw, Baier:2004tj, Gelis:2006hy, GolecBiernat:2010de, Ducloue:2017zfd}. 
	Before Sudakov evolution or a fiducial measurement, the differential cross section reads
	\begin{align}\label{eq:inclusive_cgc_born}
		\frac{d\sigma_A^{(0)}}{dy_Z\,d^2q_T} &={}\sum_q\sigma_{0,q}^{Z}
		\int d^2k_q\,d^2k_{\bar q}\, \delta^{(2)} \bigl( \boldsymbol q_T-\boldsymbol k_q-\boldsymbol k_{\bar q} \bigr) \nonumber\\[-1mm]
		&\hspace{-0.3cm}\times\Big[\mathcal B_{q/p}(x_p,\boldsymbol k_q) \mathcal B^{\rm inc}_{\bar q/A} (x_A,\boldsymbol k_{\bar q}) +\big(q\leftrightarrow\bar q\big) \Big].	
	\end{align}
	Here $\boldsymbol k_q$ and $\boldsymbol k_{\bar q}$ are the transverse momenta of the projectile quark and target antiquark entering the hard annihilation, respectively.
	$\mathcal B_{q/p}$ denotes the dilute projectile TMD, $\mathcal B^{\rm inc}_{\bar q/A}$ is the inclusive dense-target sea-quark beam function, and $\sigma_{0,q}^{Z}$ is the corresponding Born coefficient.
	Here, $q$ labels flavor and $y_Z$ is the boson rapidity in the nucleon--nucleon center-of-mass frame.
	The momentum fractions read $x_{p,A} = m_T e^{\pm y_Z}/\sqrt{s_{NN}}$ with $ m_T=\sqrt{Q^2+q_T^2}$.
	
	While the projectile transverse momentum introduces a common smearing in $p+p$ and $p+A$, the nuclear dependence resides in $\mathcal B^{\rm inc}_{\bar q/A}$.
	This target distribution provides an all-twist small-\(x\) TMD boundary condition for the subsequent Sudakov evolution, with multiple scattering and saturation encoded in its transverse momentum dependence~\cite{Balitsky:2016dgz, Altinoluk:2019wyu,Caucal:2025xxh}.
	
	At the small-\(x\) matching scale, the target sea antiquark is generated from the nuclear gluon distribution through \(g\to q\bar q\).
	The antiquark entering the hard annihilation carries $\boldsymbol k_{\bar q}=\boldsymbol\kappa_T-\boldsymbol p_T$, where $\boldsymbol\kappa_T$ is the total transverse momentum supplied to the pair, and $\boldsymbol p_T$ is the companion transverse momentum. 
	The dipole amplitude provides the distribution in $\boldsymbol\kappa_T$, and the splitting kernel controls how that momentum is shared.
	Integrating over the companion momentum yields the inclusive Drell--Yan sea-quark TMD in Eq.~\eqref{eq:inclusive_cgc_born}~\cite{Mueller:1999wm,Marquet:2009ca,Xiao:2010sa,Dominguez:2011wm,Caucal:2025xxh}. 
	For the fiducial observable, however, the acceptance measurement must be imposed before this integration, so that the companion recoil is assigned according to whether it lies inside or outside \(\mathcal C\).
	
	\emph{\textbf{Joint resummation of \(q_T\) and vector angularity.}}
	Two complementary evolutions organize the relevant logarithms~\cite{Xiao:2017yya, Caucal:2024vbv, Caucal:2025xxh}: Balitsky-Kovchegov (BK) evolution governs the small-\(x\) dependence of the target dipole distribution~\cite{Balitsky:1995ub, Kovchegov:1999yj}, whereas Collins--Soper evolution carries the matched TMD from its natural scale \(\mu_b\sim1/b\) to \(Q\simeq M_Z\)~\cite{Collins:1984kg,  Becher:2010tm, Collins:2011zzd, Catani:2015vma}.
	With tildes denoting transverse Fourier transforms, taking the CGC result as the low-scale boundary and applying Drell--Yan Sudakov evolution gives
	the inclusive spectrum
	\begin{align}\label{eq:inclusive_resummation}
		\frac{d\sigma_A^{\rm inc}}{dy_Z\,d^2q_T} ={}&\sum_q\sigma_{0,q}^{Z} \int\frac{d^2b}{(2\pi)^2}\, e^{i\boldsymbol b\cdot\boldsymbol q_T} e^{-R_{\rm DY}^{\rm pert}(b,Q)-R_{\rm NP}(b,Q)} \nonumber\\[-1mm]
		&\times \!\! \left[ \widetilde{\mathcal B}_{q/p}(x_p,b) \, \widetilde{\mathcal B}^{\rm inc}_{\bar q/A}(x_A,b) +\big(q\leftrightarrow\bar q\big) \right].
	\end{align}
	Here, $R_{\rm DY}^{\rm pert}$ is the perturbative Sudakov factor, identical to that in TMD factorization. 
	We assign the nonperturbative factor~\cite{Sun:2013hua,Sun:2014dqm,Collins:2014jpa,Wei:2020glg} \(R_{\rm NP}\) to the dilute projectile~\cite{Marquet:2019ltn}.
	The target dependence resides in \(\widetilde{\mathcal B}^{\rm inc}_{\bar q/A}\), while the perturbative Sudakov factor resums the hard-scale logarithms of \(Qb\). 
	Since \(Q\simeq M_Z\gg Q_s\), Sudakov broadening largely obscures the target-induced broadening in the inclusive \(q_T\) spectrum.
	
	In addition to the total recoil encoded in \(\boldsymbol q_T\), the fiducial measurement determines the hadronic recoil recorded inside \(\mathcal C\).
	Implementing the fiducial measurement introduced above, and excluding the
	reconstructed boson from the hadronic sum, gives
	\begin{equation}
		\boldsymbol l_T=\sum_{i\in\mathcal C}\boldsymbol p_{Ti},
		\qquad \boldsymbol k_T=\boldsymbol q_T+\boldsymbol l_T .
		\label{eq:vector_sum}
	\end{equation}
	Here, \(l_T=|\boldsymbol l_T|\) is the vector-angularity variable.
	The measurement in Eq.~\eqref{eq:vector_sum} is an inclusive hadronic sum,
	rather than a tag on the companion quark.
	At the CGC matching scale, however, the fiducial measurement must be applied to the target pair density before the companion momentum is integrated out.
	The active antiquark carries $\boldsymbol k_{\bar q}=\boldsymbol\kappa_T-\boldsymbol p_T$.  
	The target-side contribution to \(\boldsymbol k_T\) depends on whether the companion is recorded. 
	If it lies outside \(\mathcal C\), this contribution is \(\boldsymbol k_{\bar q}=\boldsymbol\kappa_T-\boldsymbol p_T\). 
	If it lies inside, \(\boldsymbol p_T\) enters \(\boldsymbol l_T\), so that the target-side contribution becomes \(\boldsymbol k_{\bar q}+\boldsymbol p_T=\boldsymbol\kappa_T\).
	
	This motivates us to define the acceptance-dependent target beam function $\widetilde{\mathcal B}^{\mathcal C}_{\bar q/A} (x_A,\boldsymbol b_k,\boldsymbol b_l)$, where $\boldsymbol b_k$ and $\boldsymbol b_l$ are conjugate to the residual momentum and the fiducial recoil, respectively. 
	Let $\mathcal P_{\bar q q/A} (x_A; \xi,\boldsymbol\kappa_T,\boldsymbol p_T)$ denote the CGC pair density, and let $\chi_{\mathcal C}\equiv \chi_{\mathcal C}(\eta_{\rm comp}^{\rm lab})$ select a companion inside the detector coverage. 
	Without the acceptance measurement, integrating the same pair density yields the inclusive target boundary, $\widetilde{\mathcal B}^{\rm inc}_{\bar q/A}(x_A,\boldsymbol b)=\int_{x_A}^{1}d\xi\int d^2\kappa_T\,d^2p_T\, \mathcal P_{\bar q q/A} e^{-i\boldsymbol b\cdot (\boldsymbol\kappa_T-\boldsymbol p_T)}$.
	Applying the acceptance before the integration instead gives
	\begin{align}
		\left.\widetilde{\mathcal B}^{\mathcal C}_{\bar q/A} (x_A,\boldsymbol b_k,\boldsymbol b_l)\right|_{\rm CGC}
		={}\int_{x_A}^{1}d\xi\int d^2\kappa_T\,d^2p_T\, \mathcal P_{\bar q q/A} \nonumber\\[-1mm]
		\times\Big[ \big(1-\chi_{\mathcal C}\big) e^{-i\boldsymbol b_k\cdot (\boldsymbol\kappa_T-\boldsymbol p_T)}
		+\chi_{\mathcal C} e^{-i\boldsymbol b_k\cdot\boldsymbol\kappa_T -i\boldsymbol b_l\cdot\boldsymbol p_T} \Big].
		\label{eq:target_beam_matching}
	\end{align}
	The two terms correspond to a companion outside and inside $\mathcal C$, respectively.  
	In the second term, $\boldsymbol b_k$ is conjugate directly to the total nuclear kick $\boldsymbol\kappa_T$.

	Perturbative radiation is resolved by the same fiducial measurement.
	Recoil deposited inside $\mathcal C$ contributes to $\boldsymbol l_T$, whereas uncovered recoil remains in the residual momentum.  
	At scale $\mu$, the incoming dipole radiates soft radiation at rapidity $y_s$ with $|y_s-y_Z|<L_\mu=\ln(Q/\mu)$.  
	The overlap of this radiation interval with the detector coverage and the corresponding Sudakov factors are
	\begin{subequations}
		\label{eq:coverage_sudakovs}
		\begin{align}
			\Delta\eta_{\rm cov}(\mu;y_Z^{\rm lab})={}&\Big[ \min\!\left(\eta_{\max}^{\rm lab},y_Z^{\rm lab}+L_\mu\right) \nonumber\\[-1mm]
			&-\max\!\left(\eta_{\min}^{\rm lab},y_Z^{\rm lab}-L_\mu\right) \Big]_+ , \label{eq:covered_rapidity}\\
			R_{\rm cov}^{\rm pert}(b_l) ={}\frac{C_F}{\pi}&\int_{\mu_{b_l}^2}^{Q^2} \frac{d\mu^2}{\mu^2}\, 	\alpha_s(\mu)\,\Delta\eta_{\rm cov}(\mu),
			\label{eq:covered_radiator}\\
			R_{\rm rest}^{\rm pert}(b_k) ={}\frac{C_F}{\pi}&\int_{\mu_{b_k}^2}^{Q^2} \frac{d\mu^2}{\mu^2}\, \alpha_s(\mu) \left[ 2L_\mu- \Delta\eta_{\rm cov}(\mu)-\frac32 \right].
			\label{eq:residual_radiator}
		\end{align}
	\end{subequations}
	Here, $y_Z^{\rm lab}$ denotes the boson rapidity in the lab frame. The Sudakov factor $R_{\rm cov}^{\rm pert}$ describes the covered radiative recoil conjugate to $\boldsymbol b_l$, whereas $R_{\rm rest}^{\rm pert}$ collects the uncovered contribution and the finite beam term at $\boldsymbol b_k$.
	The nonperturbative projectile Sudakov factor belongs to the same residual sector, so, suppressing the common $Q$ and $y_Z^{\rm lab}$ arguments, $R_{\rm rest}=R_{\rm rest}^{\rm pert}+R_{\rm NP}$.
	We present the derivation of these Sudakov factors and more details on the pair kernel of \cref{eq:target_beam_matching} in the Supplemental Material~\cite{SupplementalMaterial}.

	Combining the acceptance-dependent target boundary with the two Sudakov factors above, the joint recoil distribution factorizes as
	\begin{align}
		\frac{d\sigma_A}{dy_Z\,d^2q_T\,d^2l_T} ={}&\sum_q\sigma_{0,q}^{Z}\int\frac{d^2b_k\,d^2b_l}{(2\pi)^4}\, e^{i\boldsymbol b_k\cdot\boldsymbol k_T +i\boldsymbol b_l\cdot\boldsymbol l_T}  \nonumber\\[-1mm]
		&\times\Big[ \widetilde{\mathcal B}_{q/p}(x_p,\boldsymbol{b}_k) \widetilde{\mathcal B}^{\mathcal C}_{\bar q/A} (x_A,\boldsymbol b_k,\boldsymbol b_l) \nonumber\\[-1mm]
		&+\big(q\leftrightarrow\bar q\big) \Big]\,e^{-R_{\rm rest}(b_k)-R_{\rm cov}^{\rm pert}(b_l)}.
		\label{eq:joint_cross_section}
	\end{align}
	The spectrum subject to a cumulative vector angularity requirement is then given by
	\begin{equation}
		\left.\frac{d\sigma_A}{dy_Z\,d^2q_T}\right|_{l_T<\tau_0} \equiv \int_{l_T<\tau_0}d^2l_T\, \frac{d\sigma_A}{dy_Z\,d^2q_T\,d^2l_T}.
		\label{eq:vector_jettiness_cumulant}
	\end{equation}
	Unconstrained integration over $\boldsymbol l_T$ enforces $\boldsymbol b_l=-\boldsymbol b_k \equiv -\boldsymbol b$. 
	An upper cut $\tau_0$ on $l_T$ is only applied after the measured vector is formed; its $\tau_0\to\infty$ limit reproduces Eq.~\eqref{eq:inclusive_resummation}. 
	The boundary condition thus simplifies to $\widetilde{\mathcal B}^{\mathcal C}_{\bar q/A} (x_A,\boldsymbol b,-\boldsymbol b) = \widetilde{\mathcal B}^{\rm inc}_{\bar q/A}(x_A,b)$.
	Using $R_{\rm cov}^{\rm pert}(b)+R_{\rm rest}(b) =R_{\rm DY}^{\rm pert}(b)+R_{\rm NP}(b)$, this integration analytically recovers the inclusive form in Eq.~\eqref{eq:inclusive_resummation}. 
	Unlike dijet or dihadron acoplanarity, $q\bar q\to Z^0$ has no colored hard final-state particle and therefore introduces neither a final-state Sudakov factor nor fragmentation smearing~\cite{Deak:2018obv, Goncalves:2020tvh, Benic:2022ixp, Taels:2022tza, Yang:2022qgk, Ganguli:2023joy, Taels:2023czt, Caucal:2023fsf, Caucal:2024nsb, Bandeira:2024jjl}. 
	At leading power, the remaining hard-scale radiative broadening of $\boldsymbol k_T$ comes from initial-state recoil outside $\mathcal C$, while target-side energy flow is encoded in $\widetilde{\mathcal B}^{\mathcal C}_{\bar q/A}$.
	
	Once the soft-radiation interval contains the full detector coverage, the covered region removes a fixed rapidity length	$\Delta\eta_{\rm cov} =\eta_{\max}^{\rm lab}-\eta_{\min}^{\rm lab}\equiv\Delta\eta_{\mathcal C}^{\rm lab}$ from the cusp phase space contributing to the residual recoil. 
	The remaining phase space that contributes to the Sudakov evolution, i.e. the sum of the two uncovered regions, has rapidity extent
	\begin{equation}
		2L_\mu-\Delta\eta_{\mathcal C}^{\rm lab} \equiv \ln\frac{Q_{\rm eff}^2}{\mu^2},
		\qquad Q_{\rm eff}= Qe^{-\Delta\eta_{\mathcal C}^{\rm lab}/2}.
		\label{eq:effective_sudakov_scale}
	\end{equation}
	For $\Delta\eta_{\mathcal C}^{\rm lab}=5$, we obtain $Q_{\rm eff} \simeq 7.5$ GeV. 
	Thus $Q_{\rm eff}$ characterizes the double-logarithmic radiative phase space that remains in the residual momentum: increasing the coverage lowers this scale and reduces the perturbative broadening of $\boldsymbol k_T$. 
	When the boson is displaced from the center of the coverage, the two boundaries become active at different values of $\mu$; their exact turn-on is retained in Eq.~\eqref{eq:coverage_sudakovs}, rather than replacing the full Sudakov factor by a single effective scale.  
	The hard matching scale remains $Q\simeq M_Z$.

	\emph{\textbf{Phenomenology.}}
	The joint factorization predicts a correlated response to the target and the fiducial coverage: enlarging \(\mathcal C\) reduces the hard-scale radiative broadening of \(\boldsymbol k_T\), whereas the target dependence remains encoded in \(\widetilde{\mathcal B}^{\mathcal C}_{\bar q/A}\).
	We evolve the target dipole amplitude with the running-coupling BK equation~\cite{Albacete:2010sy}, starting from the MV-model initial condition~\cite{McLerran:1993ni,McLerran:1993ka}.
	We evaluate the joint resummed distributions using the one-loop perturbative Sudakov factor at $\sqrt{s_{NN}}=8.16~\mathrm{TeV}$ and $y_Z=1.5$. Further details on the kinematics, matching, and resummation ingredients are given in the Supplemental Material~\cite{SupplementalMaterial}.
	The pp, pA2, and pA3 (A = Pb) calculations use common hard and projectile inputs, while their initial saturation scales satisfy \(Q_{s0,\,\mathrm{pA2}}^2=2Q_{s0,\,p}^2\) and \(Q_{s0,\,\mathrm{pA3}}^2=3Q_{s0,\,p}^2\), with $Q_{s0,\,p}^2=0.2$ GeV$^2$.  
	The two nuclear inputs serve as benchmarks for weaker and stronger saturation.

	\begin{figure}[t]
		\centering
		\includegraphics[width=0.9\columnwidth]
		{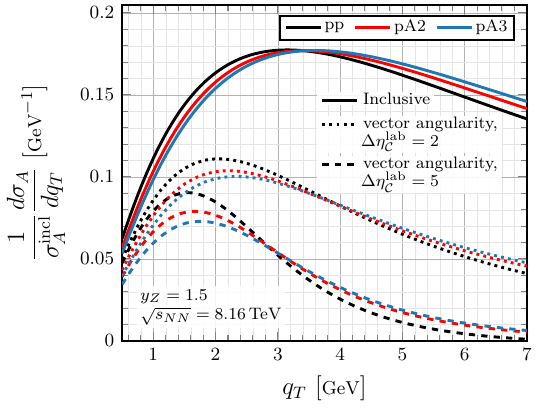}
		\caption{Normalized $Z^0$ transverse-momentum densities at $y_Z=1.5$ and $\sqrt{s_{NN}}=8.16~\mathrm{TeV}$. 
			Solid curves are inclusive, while dashed and dotted curves impose \(l_T<2~\mathrm{GeV}\) for $\Delta\eta_{\mathcal C}^{\rm lab}=5$ and $\Delta\eta_{\mathcal C}^{\rm lab}=2$, respectively. 
			For each target, all curves are divided by that target's inclusive integral over $0.5<q_T<7~\mathrm{GeV}$. 
			Consequently, the integral of a dashed or dotted curve over this interval gives the corresponding selected cross-section fraction.}
		\label{fig:qt_spectra}
	\end{figure}
	
	Fig.~\ref{fig:qt_spectra} compares the inclusive \(q_T\) distributions with those selected by the cumulative requirement \(l_T<2~\mathrm{GeV}\). 
	For each target, the inclusive and selected curves are divided by the same inclusive integral over the displayed $q_T$ range \(0.5<q_T<7~\mathrm{GeV}\).  
	The inclusive distributions are nearly insensitive to the target input because evolution to \(M_Z\) largely obscures the transverse momentum generated at the target scale.  
	The selected spectra exhibit a more pronounced ordering: increasing \(Q_s\) depletes the distribution around its maximum and shifts weight toward larger \(q_T\).  
	The selection retains about \(25\%\) of the inclusive cross section for \(\Delta\eta_{\mathcal C}^{\rm lab}=5\) and about \(50\%\) for \(\Delta\eta_{\mathcal C}^{\rm lab}=2\).

	We next consider the recoil-angle distribution obtained from the same joint cross section.
	While vector angularity constrains the magnitude $l_T$, the angular correlation between the boson and the fiducial recoil offers an even more selective probe. 
	A soft gluon emitted inside $\mathcal C$ shifts $\boldsymbol q_T$ and $\boldsymbol l_T$ in exactly opposite directions, leaving the residual momentum $\boldsymbol k_T=\boldsymbol q_T+\boldsymbol l_T$ invariant. 
	At fixed $q_T$ and $l_T$, the angle probes the residual convolution of projectile transverse momentum, the nuclear beam function, and uncovered radiation.  
	The cancellation becomes more effective as the rapidity coverage grows. Defining $\Delta\phi_{Zl}$ as the azimuthal angle between $\boldsymbol{l}_T$ and the transverse momentum of the $Z^0$-boson, we obtain $k_T\simeq q_T(\pi-\Delta\phi_{Zl})$ when $q_T\simeq l_T$.
	
\begin{figure}[t]
		\centering
		\includegraphics[width=0.9\columnwidth]
		{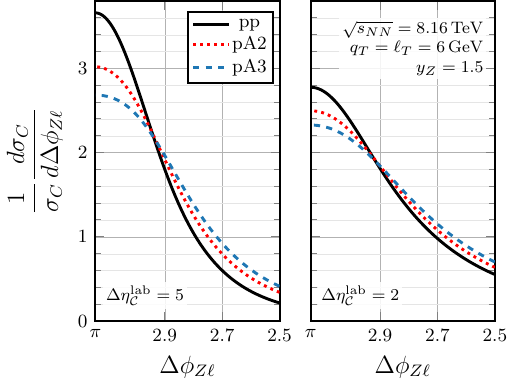}
		\caption{Recoil-angle distributions at $q_T=l_T=6~\mathrm{GeV}$, $y_Z=1.5$, and $\sqrt{s_{NN}}=8.16~\mathrm{TeV}$.  
			The left and right panels use the symmetric laboratory-frame coverages $|\eta^{\rm lab}|<2.5$ and $|\eta^{\rm lab}|<1$, corresponding to $\Delta\eta_{\mathcal C}^{\rm lab}=5$ and $2$ and to $Q_{\rm eff}\simeq7.50$ and $33.6~\mathrm{GeV}$, respectively. 
			Each curve is normalized over $2.5<\Delta\phi_{Zl}<\pi$.}
		\label{fig:recoil_angle_distribution}
	\end{figure}

	For the symmetric configuration $q_T=l_T=6~\mathrm{GeV}$, Fig.~\ref{fig:recoil_angle_distribution} shows that increasing $Q_s$ shifts probability away from the back-to-back limit, $\Delta\phi_{Zl}=\pi$.  
	The separation between the nuclear targets and the $pp$ baseline near this limit is considerably larger for $\Delta\eta_{\mathcal C}^{\rm lab}=5$ than for $\Delta\eta_{\mathcal C}^{\rm lab}=2$.  
	Enlarging the coverage lowers $Q_{\rm eff}$ from $33.6$ to $7.50~\mathrm{GeV}$, such that a larger fraction of the perturbative recoil is reconstructed in $\boldsymbol l_T$ and cancels from $\boldsymbol k_T=\boldsymbol q_T+\boldsymbol l_T$. 
	The reduced Sudakov broadening consequently makes the target-induced angular shift more visible.
	
\begin{figure}[t]
		\centering
		\includegraphics[width=\columnwidth]
		{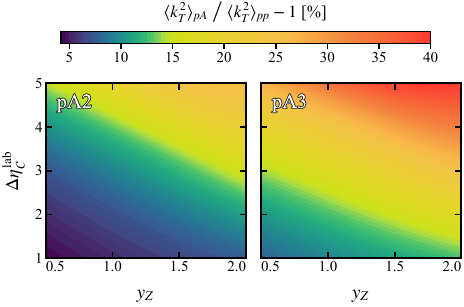}
		\caption{Relative nuclear broadening of the residual transverse
			momentum for the pA2 (left) and pA3 (right) benchmarks at
			$q_T=l_T=6~\mathrm{GeV}$ and $\sqrt{s_{NN}}=8.16~\mathrm{TeV}$.  The color scale displays $\langle k_T^2\rangle_{pA}/\langle k_T^2\rangle_{pp}-1$.  
			At each point, residual transverse-momentum broadening $\langle k_T^2\rangle$ is evaluated from the recoil-angle distribution normalized over $2.5<\Delta\phi_{Zl}<\pi$.  
			The coverage is the symmetric laboratory-frame interval $|\eta^{\rm lab}|<\Delta\eta_{\mathcal C}^{\rm lab}/2$.}
		\label{fig:coverage_heatmap}
	\end{figure}

	This enhancement with coverage persists over the rapidity--coverage plane.  
	Fig.~\ref{fig:coverage_heatmap} shows the nuclear enhancement of the residual transverse-momentum broadening, $\langle k_T^2\rangle_{pA}/\langle k_T^2\rangle_{pp}-1$. 
	The enhancement follows the expected saturation pattern: it is stronger for pA3, which has the larger saturation scale, and becomes more pronounced toward forward rapidity as the probed target momentum fraction $x_A$ decreases.
	At fixed target and rapidity, the enhancement also grows systematically with $\Delta\eta_{\mathcal C}^{\rm lab}$, by the same mechanism observed in \cref{fig:recoil_angle_distribution}.  
	
	\emph{\textbf{Conclusions.}}
	We have introduced a fiducial vector-recoil observable that separates the transverse momentum recorded within a chosen rapidity interval from the residual recoil. 
	In forward $Z^0$ production, the vector sum of the boson momentum and the measured hadronic recoil cancels the in-acceptance radiative contribution at leading power, while retaining sensitivity to the target momentum transfer together with projectile smearing and radiation outside the coverage. 
	For $|\eta^{\rm lab}|<2.5$, the uncovered Sudakov phase space is characterized by $Q_{\rm eff}\simeq7.5~\mathrm{GeV}$, substantially below $M_Z$.
	
	Our calculation, which combines a CGC description of the small-$x$ target with SCET resummation, predicts a correlated response to the target and detector geometry.
	Increasing $Q_s$ broadens the residual-momentum distribution and weakens recoil alignment, whereas enlarging the coverage suppresses the radiative background and makes the proton-nucleus separation more visible. 
	The combined dependence on target input, rapidity, and acceptance therefore provides a discriminating test of whether the observed broadening originates from nonlinear small-$x$ dynamics rather than from hard-scale radiation alone.

	\emph{\textbf{Acknowledgments.}} 
	This work has been supported by the National Natural Science Foundation of China under Grants No.~12175118 (J.Z.), No.~12321005 (J.Z.), No.~12275052 (W. L. and D.Y.S.), No.~12547102 (D.Y.S.),  No.~12147101 (D.Y.S.) and No.~12405156 (S.Y.W.). 
	W.~L. is also supported by the China Postdoctoral Science Foundation under grant No.~2025M783369. 
	D.Y.S. is also supported by the Innovation Program for Quantum Science and Technology under Grant No. 2024ZD0300101.
	
	\bibliographystyle{apsrev4-2}
	\bibliography{ref}

\begin{thebibliography}{114}%
\makeatletter
\providecommand \@ifxundefined [1]{%
 \@ifx{#1\undefined}
}%
\providecommand \@ifnum [1]{%
 \ifnum #1\expandafter \@firstoftwo
 \else \expandafter \@secondoftwo
 \fi
}%
\providecommand \@ifx [1]{%
 \ifx #1\expandafter \@firstoftwo
 \else \expandafter \@secondoftwo
 \fi
}%
\providecommand \natexlab [1]{#1}%
\providecommand \enquote  [1]{``#1''}%
\providecommand \bibnamefont  [1]{#1}%
\providecommand \bibfnamefont [1]{#1}%
\providecommand \citenamefont [1]{#1}%
\providecommand \href@noop [0]{\@secondoftwo}%
\providecommand \href [0]{\begingroup \@sanitize@url \@href}%
\providecommand \@href[1]{\@@startlink{#1}\@@href}%
\providecommand \@@href[1]{\endgroup#1\@@endlink}%
\providecommand \@sanitize@url [0]{\catcode `\\12\catcode `\$12\catcode
  `\&12\catcode `\#12\catcode `\^12\catcode `\_12\catcode `\%12\relax}%
\providecommand \@@startlink[1]{}%
\providecommand \@@endlink[0]{}%
\providecommand \url  [0]{\begingroup\@sanitize@url \@url }%
\providecommand \@url [1]{\endgroup\@href {#1}{\urlprefix }}%
\providecommand \urlprefix  [0]{URL }%
\providecommand \Eprint [0]{\href }%
\providecommand \doibase [0]{https://doi.org/}%
\providecommand \selectlanguage [0]{\@gobble}%
\providecommand \bibinfo  [0]{\@secondoftwo}%
\providecommand \bibfield  [0]{\@secondoftwo}%
\providecommand \translation [1]{[#1]}%
\providecommand \BibitemOpen [0]{}%
\providecommand \bibitemStop [0]{}%
\providecommand \bibitemNoStop [0]{.\EOS\space}%
\providecommand \EOS [0]{\spacefactor3000\relax}%
\providecommand \BibitemShut  [1]{\csname bibitem#1\endcsname}%
\let\auto@bib@innerbib\@empty
\bibitem [{\citenamefont {Gribov}\ \emph {et~al.}(1983)\citenamefont {Gribov},
  \citenamefont {Levin},\ and\ \citenamefont {Ryskin}}]{Gribov:1984tu}%
  \BibitemOpen
  \bibfield  {author} {\bibinfo {author} {\bibfnamefont {L.~V.}\ \bibnamefont
  {Gribov}}, \bibinfo {author} {\bibfnamefont {E.~M.}\ \bibnamefont {Levin}},\
  and\ \bibinfo {author} {\bibfnamefont {M.~G.}\ \bibnamefont {Ryskin}},\
  }\href {https://doi.org/10.1016/0370-1573(83)90022-4} {\bibfield  {journal}
  {\bibinfo  {journal} {Phys. Rept.}\ }\textbf {\bibinfo {volume} {100}},\
  \bibinfo {pages} {1} (\bibinfo {year} {1983})}\BibitemShut {NoStop}%
\bibitem [{\citenamefont {Mueller}\ and\ \citenamefont
  {Qiu}(1986)}]{Mueller:1985wy}%
  \BibitemOpen
  \bibfield  {author} {\bibinfo {author} {\bibfnamefont {A.~H.}\ \bibnamefont
  {Mueller}}\ and\ \bibinfo {author} {\bibfnamefont {J.-w.}\ \bibnamefont
  {Qiu}},\ }\href {https://doi.org/10.1016/0550-3213(86)90164-1} {\bibfield
  {journal} {\bibinfo  {journal} {Nucl. Phys. B}\ }\textbf {\bibinfo {volume}
  {268}},\ \bibinfo {pages} {427} (\bibinfo {year} {1986})}\BibitemShut
  {NoStop}%
\bibitem [{\citenamefont {McLerran}\ and\ \citenamefont
  {Venugopalan}(1994{\natexlab{a}})}]{McLerran:1993ni}%
  \BibitemOpen
  \bibfield  {author} {\bibinfo {author} {\bibfnamefont {L.~D.}\ \bibnamefont
  {McLerran}}\ and\ \bibinfo {author} {\bibfnamefont {R.}~\bibnamefont
  {Venugopalan}},\ }\href {https://doi.org/10.1103/PhysRevD.49.2233} {\bibfield
   {journal} {\bibinfo  {journal} {Phys. Rev. D}\ }\textbf {\bibinfo {volume}
  {49}},\ \bibinfo {pages} {2233} (\bibinfo {year} {1994}{\natexlab{a}})},\
  \Eprint {https://arxiv.org/abs/hep-ph/9309289} {arXiv:hep-ph/9309289}
  \BibitemShut {NoStop}%
\bibitem [{\citenamefont {McLerran}\ and\ \citenamefont
  {Venugopalan}(1994{\natexlab{b}})}]{McLerran:1993ka}%
  \BibitemOpen
  \bibfield  {author} {\bibinfo {author} {\bibfnamefont {L.~D.}\ \bibnamefont
  {McLerran}}\ and\ \bibinfo {author} {\bibfnamefont {R.}~\bibnamefont
  {Venugopalan}},\ }\href {https://doi.org/10.1103/PhysRevD.49.3352} {\bibfield
   {journal} {\bibinfo  {journal} {Phys. Rev. D}\ }\textbf {\bibinfo {volume}
  {49}},\ \bibinfo {pages} {3352} (\bibinfo {year} {1994}{\natexlab{b}})},\
  \Eprint {https://arxiv.org/abs/hep-ph/9311205} {arXiv:hep-ph/9311205}
  \BibitemShut {NoStop}%
\bibitem [{\citenamefont {Mueller}(1999)}]{Mueller:1999wm}%
  \BibitemOpen
  \bibfield  {author} {\bibinfo {author} {\bibfnamefont {A.~H.}\ \bibnamefont
  {Mueller}},\ }\href {https://doi.org/10.1016/S0550-3213(99)00394-6}
  {\bibfield  {journal} {\bibinfo  {journal} {Nucl. Phys. B}\ }\textbf
  {\bibinfo {volume} {558}},\ \bibinfo {pages} {285} (\bibinfo {year}
  {1999})},\ \Eprint {https://arxiv.org/abs/hep-ph/9904404}
  {arXiv:hep-ph/9904404} \BibitemShut {NoStop}%
\bibitem [{\citenamefont {Gelis}\ \emph {et~al.}(2010)\citenamefont {Gelis},
  \citenamefont {Iancu}, \citenamefont {Jalilian-Marian},\ and\ \citenamefont
  {Venugopalan}}]{Gelis:2010nm}%
  \BibitemOpen
  \bibfield  {author} {\bibinfo {author} {\bibfnamefont {F.}~\bibnamefont
  {Gelis}}, \bibinfo {author} {\bibfnamefont {E.}~\bibnamefont {Iancu}},
  \bibinfo {author} {\bibfnamefont {J.}~\bibnamefont {Jalilian-Marian}},\ and\
  \bibinfo {author} {\bibfnamefont {R.}~\bibnamefont {Venugopalan}},\ }\href
  {https://doi.org/10.1146/annurev.nucl.010909.083629} {\bibfield  {journal}
  {\bibinfo  {journal} {Ann. Rev. Nucl. Part. Sci.}\ }\textbf {\bibinfo
  {volume} {60}},\ \bibinfo {pages} {463} (\bibinfo {year} {2010})},\ \Eprint
  {https://arxiv.org/abs/1002.0333} {arXiv:1002.0333 [hep-ph]} \BibitemShut
  {NoStop}%
\bibitem [{\citenamefont {Sch\"afer}\ and\ \citenamefont
  {Zhou}(2013)}]{Schafer:2013mza}%
  \BibitemOpen
  \bibfield  {author} {\bibinfo {author} {\bibfnamefont {A.}~\bibnamefont
  {Sch\"afer}}\ and\ \bibinfo {author} {\bibfnamefont {J.}~\bibnamefont
  {Zhou}},\ }\href {https://doi.org/10.1103/PhysRevD.88.074012} {\bibfield
  {journal} {\bibinfo  {journal} {Phys. Rev. D}\ }\textbf {\bibinfo {volume}
  {88}},\ \bibinfo {pages} {074012} (\bibinfo {year} {2013})},\ \Eprint
  {https://arxiv.org/abs/1305.5042} {arXiv:1305.5042 [hep-ph]} \BibitemShut
  {NoStop}%
\bibitem [{\citenamefont {Albacete}\ \emph {et~al.}(2019)\citenamefont
  {Albacete}, \citenamefont {Giacalone}, \citenamefont {Marquet},\ and\
  \citenamefont {Matas}}]{Albacete:2018ruq}%
  \BibitemOpen
  \bibfield  {author} {\bibinfo {author} {\bibfnamefont {J.~L.}\ \bibnamefont
  {Albacete}}, \bibinfo {author} {\bibfnamefont {G.}~\bibnamefont {Giacalone}},
  \bibinfo {author} {\bibfnamefont {C.}~\bibnamefont {Marquet}},\ and\ \bibinfo
  {author} {\bibfnamefont {M.}~\bibnamefont {Matas}},\ }\href
  {https://doi.org/10.1103/PhysRevD.99.014002} {\bibfield  {journal} {\bibinfo
  {journal} {Phys. Rev. D}\ }\textbf {\bibinfo {volume} {99}},\ \bibinfo
  {pages} {014002} (\bibinfo {year} {2019})},\ \Eprint
  {https://arxiv.org/abs/1805.05711} {arXiv:1805.05711 [hep-ph]} \BibitemShut
  {NoStop}%
\bibitem [{\citenamefont {Adare}\ \emph {et~al.}(2011)\citenamefont {Adare}
  \emph {et~al.}}]{Adare:2011sc}%
  \BibitemOpen
  \bibfield  {author} {\bibinfo {author} {\bibfnamefont {A.}~\bibnamefont
  {Adare}} \emph {et~al.} (\bibinfo {collaboration} {PHENIX}),\ }\href
  {https://doi.org/10.1103/PhysRevLett.107.172301} {\bibfield  {journal}
  {\bibinfo  {journal} {Phys. Rev. Lett.}\ }\textbf {\bibinfo {volume} {107}},\
  \bibinfo {pages} {172301} (\bibinfo {year} {2011})},\ \Eprint
  {https://arxiv.org/abs/1105.5112} {arXiv:1105.5112 [nucl-ex]} \BibitemShut
  {NoStop}%
\bibitem [{\citenamefont {Aaboud}\ \emph {et~al.}(2019)\citenamefont {Aaboud}
  \emph {et~al.}}]{Aaboud:2019oop}%
  \BibitemOpen
  \bibfield  {author} {\bibinfo {author} {\bibfnamefont {M.}~\bibnamefont
  {Aaboud}} \emph {et~al.} (\bibinfo {collaboration} {ATLAS}),\ }\href
  {https://doi.org/10.1103/PhysRevC.100.034903} {\bibfield  {journal} {\bibinfo
   {journal} {Phys. Rev. C}\ }\textbf {\bibinfo {volume} {100}},\ \bibinfo
  {pages} {034903} (\bibinfo {year} {2019})},\ \Eprint
  {https://arxiv.org/abs/1901.10440} {arXiv:1901.10440 [nucl-ex]} \BibitemShut
  {NoStop}%
\bibitem [{\citenamefont {Albacete}\ and\ \citenamefont
  {Marquet}(2010)}]{Albacete:2010pg}%
  \BibitemOpen
  \bibfield  {author} {\bibinfo {author} {\bibfnamefont {J.~L.}\ \bibnamefont
  {Albacete}}\ and\ \bibinfo {author} {\bibfnamefont {C.}~\bibnamefont
  {Marquet}},\ }\href {https://doi.org/10.1103/PhysRevLett.105.162301}
  {\bibfield  {journal} {\bibinfo  {journal} {Phys. Rev. Lett.}\ }\textbf
  {\bibinfo {volume} {105}},\ \bibinfo {pages} {162301} (\bibinfo {year}
  {2010})},\ \Eprint {https://arxiv.org/abs/1005.4065} {arXiv:1005.4065
  [hep-ph]} \BibitemShut {NoStop}%
\bibitem [{\citenamefont {Lappi}\ and\ \citenamefont
  {Mantysaari}(2013)}]{Lappi:2012nh}%
  \BibitemOpen
  \bibfield  {author} {\bibinfo {author} {\bibfnamefont {T.}~\bibnamefont
  {Lappi}}\ and\ \bibinfo {author} {\bibfnamefont {H.}~\bibnamefont
  {Mantysaari}},\ }\href {https://doi.org/10.1016/j.nuclphysa.2013.03.017}
  {\bibfield  {journal} {\bibinfo  {journal} {Nucl. Phys. A}\ }\textbf
  {\bibinfo {volume} {908}},\ \bibinfo {pages} {51} (\bibinfo {year} {2013})},\
  \Eprint {https://arxiv.org/abs/1209.2853} {arXiv:1209.2853 [hep-ph]}
  \BibitemShut {NoStop}%
\bibitem [{\citenamefont {Akcakaya}\ \emph {et~al.}(2013)\citenamefont
  {Akcakaya}, \citenamefont {Sch\"afer},\ and\ \citenamefont
  {Zhou}}]{Akcakaya:2012si}%
  \BibitemOpen
  \bibfield  {author} {\bibinfo {author} {\bibfnamefont {E.}~\bibnamefont
  {Akcakaya}}, \bibinfo {author} {\bibfnamefont {A.}~\bibnamefont
  {Sch\"afer}},\ and\ \bibinfo {author} {\bibfnamefont {J.}~\bibnamefont
  {Zhou}},\ }\href {https://doi.org/10.1103/PhysRevD.87.054010} {\bibfield
  {journal} {\bibinfo  {journal} {Phys. Rev. D}\ }\textbf {\bibinfo {volume}
  {87}},\ \bibinfo {pages} {054010} (\bibinfo {year} {2013})},\ \Eprint
  {https://arxiv.org/abs/1208.4965} {arXiv:1208.4965 [hep-ph]} \BibitemShut
  {NoStop}%
\bibitem [{\citenamefont {Stasto}\ \emph {et~al.}(2012)\citenamefont {Stasto},
  \citenamefont {Xiao},\ and\ \citenamefont {Zaslavsky}}]{Stasto:2012ru}%
  \BibitemOpen
  \bibfield  {author} {\bibinfo {author} {\bibfnamefont {A.}~\bibnamefont
  {Stasto}}, \bibinfo {author} {\bibfnamefont {B.-W.}\ \bibnamefont {Xiao}},\
  and\ \bibinfo {author} {\bibfnamefont {D.}~\bibnamefont {Zaslavsky}},\ }\href
  {https://doi.org/10.1103/PhysRevD.86.014009} {\bibfield  {journal} {\bibinfo
  {journal} {Phys. Rev. D}\ }\textbf {\bibinfo {volume} {86}},\ \bibinfo
  {pages} {014009} (\bibinfo {year} {2012})},\ \Eprint
  {https://arxiv.org/abs/1204.4861} {arXiv:1204.4861 [hep-ph]} \BibitemShut
  {NoStop}%
\bibitem [{\citenamefont {Kutak}\ and\ \citenamefont
  {Sapeta}(2012)}]{Kutak:2012rf}%
  \BibitemOpen
  \bibfield  {author} {\bibinfo {author} {\bibfnamefont {K.}~\bibnamefont
  {Kutak}}\ and\ \bibinfo {author} {\bibfnamefont {S.}~\bibnamefont {Sapeta}},\
  }\href {https://doi.org/10.1103/PhysRevD.86.094043} {\bibfield  {journal}
  {\bibinfo  {journal} {Phys. Rev. D}\ }\textbf {\bibinfo {volume} {86}},\
  \bibinfo {pages} {094043} (\bibinfo {year} {2012})},\ \Eprint
  {https://arxiv.org/abs/1205.5035} {arXiv:1205.5035 [hep-ph]} \BibitemShut
  {NoStop}%
\bibitem [{\citenamefont {Kotko}\ \emph {et~al.}(2015)\citenamefont {Kotko},
  \citenamefont {Kutak}, \citenamefont {Marquet}, \citenamefont {Petreska},
  \citenamefont {Sapeta},\ and\ \citenamefont {van Hameren}}]{Kotko:2015ura}%
  \BibitemOpen
  \bibfield  {author} {\bibinfo {author} {\bibfnamefont {P.}~\bibnamefont
  {Kotko}}, \bibinfo {author} {\bibfnamefont {K.}~\bibnamefont {Kutak}},
  \bibinfo {author} {\bibfnamefont {C.}~\bibnamefont {Marquet}}, \bibinfo
  {author} {\bibfnamefont {E.}~\bibnamefont {Petreska}}, \bibinfo {author}
  {\bibfnamefont {S.}~\bibnamefont {Sapeta}},\ and\ \bibinfo {author}
  {\bibfnamefont {A.}~\bibnamefont {van Hameren}},\ }\href
  {https://doi.org/10.1007/JHEP09(2015)106} {\bibfield  {journal} {\bibinfo
  {journal} {JHEP}\ }\textbf {\bibinfo {volume} {2015}}\bibfield  {number}
  {\bibinfo  {number} { (09)},\ \bibinfo {pages} {106}},\ }\Eprint
  {https://arxiv.org/abs/1503.03421} {arXiv:1503.03421 [hep-ph]} \BibitemShut
  {NoStop}%
\bibitem [{\citenamefont {van Hameren}\ \emph {et~al.}(2016)\citenamefont {van
  Hameren}, \citenamefont {Kotko}, \citenamefont {Kutak}, \citenamefont
  {Marquet}, \citenamefont {Petreska},\ and\ \citenamefont
  {Sapeta}}]{vanHameren:2016ftb}%
  \BibitemOpen
  \bibfield  {author} {\bibinfo {author} {\bibfnamefont {A.}~\bibnamefont {van
  Hameren}}, \bibinfo {author} {\bibfnamefont {P.}~\bibnamefont {Kotko}},
  \bibinfo {author} {\bibfnamefont {K.}~\bibnamefont {Kutak}}, \bibinfo
  {author} {\bibfnamefont {C.}~\bibnamefont {Marquet}}, \bibinfo {author}
  {\bibfnamefont {E.}~\bibnamefont {Petreska}},\ and\ \bibinfo {author}
  {\bibfnamefont {S.}~\bibnamefont {Sapeta}},\ }\href
  {https://doi.org/10.1007/JHEP12(2016)034} {\bibfield  {journal} {\bibinfo
  {journal} {JHEP}\ }\textbf {\bibinfo {volume} {2016}}\bibfield  {number}
  {\bibinfo  {number} { (12)},\ \bibinfo {pages} {034}},\ }\bibinfo {note}
  {[Erratum: JHEP 02, 158 (2019)]},\ \Eprint {https://arxiv.org/abs/1607.03121}
  {arXiv:1607.03121 [hep-ph]} \BibitemShut {NoStop}%
\bibitem [{\citenamefont {Stasto}\ \emph {et~al.}(2018)\citenamefont {Stasto},
  \citenamefont {Wei}, \citenamefont {Xiao},\ and\ \citenamefont
  {Yuan}}]{Stasto:2018rci}%
  \BibitemOpen
  \bibfield  {author} {\bibinfo {author} {\bibfnamefont {A.}~\bibnamefont
  {Stasto}}, \bibinfo {author} {\bibfnamefont {S.-Y.}\ \bibnamefont {Wei}},
  \bibinfo {author} {\bibfnamefont {B.-W.}\ \bibnamefont {Xiao}},\ and\
  \bibinfo {author} {\bibfnamefont {F.}~\bibnamefont {Yuan}},\ }\href
  {https://doi.org/10.1016/j.physletb.2018.08.011} {\bibfield  {journal}
  {\bibinfo  {journal} {Phys. Lett. B}\ }\textbf {\bibinfo {volume} {784}},\
  \bibinfo {pages} {301} (\bibinfo {year} {2018})},\ \Eprint
  {https://arxiv.org/abs/1805.05712} {arXiv:1805.05712 [hep-ph]} \BibitemShut
  {NoStop}%
\bibitem [{\citenamefont {van Hameren}\ \emph {et~al.}(2019)\citenamefont {van
  Hameren}, \citenamefont {Kotko}, \citenamefont {Kutak},\ and\ \citenamefont
  {Sapeta}}]{vanHameren:2019ysa}%
  \BibitemOpen
  \bibfield  {author} {\bibinfo {author} {\bibfnamefont {A.}~\bibnamefont {van
  Hameren}}, \bibinfo {author} {\bibfnamefont {P.}~\bibnamefont {Kotko}},
  \bibinfo {author} {\bibfnamefont {K.}~\bibnamefont {Kutak}},\ and\ \bibinfo
  {author} {\bibfnamefont {S.}~\bibnamefont {Sapeta}},\ }\href
  {https://doi.org/10.1016/j.physletb.2019.06.055} {\bibfield  {journal}
  {\bibinfo  {journal} {Phys. Lett. B}\ }\textbf {\bibinfo {volume} {795}},\
  \bibinfo {pages} {511} (\bibinfo {year} {2019})},\ \Eprint
  {https://arxiv.org/abs/1903.01361} {arXiv:1903.01361 [hep-ph]} \BibitemShut
  {NoStop}%
\bibitem [{\citenamefont {Al-Mashad}\ \emph {et~al.}(2022)\citenamefont
  {Al-Mashad}, \citenamefont {van Hameren}, \citenamefont {Kakkad},
  \citenamefont {Kotko}, \citenamefont {Kutak}, \citenamefont {van Mechelen},\
  and\ \citenamefont {Sapeta}}]{Al-Mashad:2022zbq}%
  \BibitemOpen
  \bibfield  {author} {\bibinfo {author} {\bibfnamefont {M.~A.}\ \bibnamefont
  {Al-Mashad}}, \bibinfo {author} {\bibfnamefont {A.}~\bibnamefont {van
  Hameren}}, \bibinfo {author} {\bibfnamefont {H.}~\bibnamefont {Kakkad}},
  \bibinfo {author} {\bibfnamefont {P.}~\bibnamefont {Kotko}}, \bibinfo
  {author} {\bibfnamefont {K.}~\bibnamefont {Kutak}}, \bibinfo {author}
  {\bibfnamefont {P.}~\bibnamefont {van Mechelen}},\ and\ \bibinfo {author}
  {\bibfnamefont {S.}~\bibnamefont {Sapeta}},\ }\href
  {https://doi.org/10.1007/JHEP12(2022)131} {\bibfield  {journal} {\bibinfo
  {journal} {JHEP}\ }\textbf {\bibinfo {volume} {2022}}\bibfield  {number}
  {\bibinfo  {number} { (12)},\ \bibinfo {pages} {131}},\ }\Eprint
  {https://arxiv.org/abs/2210.06613} {arXiv:2210.06613 [hep-ph]} \BibitemShut
  {NoStop}%
\bibitem [{\citenamefont {van Hameren}\ \emph {et~al.}(2023)\citenamefont {van
  Hameren}, \citenamefont {Kakkad}, \citenamefont {Kotko}, \citenamefont
  {Kutak},\ and\ \citenamefont {Sapeta}}]{vanHameren:2023oiq}%
  \BibitemOpen
  \bibfield  {author} {\bibinfo {author} {\bibfnamefont {A.}~\bibnamefont {van
  Hameren}}, \bibinfo {author} {\bibfnamefont {H.}~\bibnamefont {Kakkad}},
  \bibinfo {author} {\bibfnamefont {P.}~\bibnamefont {Kotko}}, \bibinfo
  {author} {\bibfnamefont {K.}~\bibnamefont {Kutak}},\ and\ \bibinfo {author}
  {\bibfnamefont {S.}~\bibnamefont {Sapeta}},\ }\href
  {https://doi.org/10.1140/epjc/s10052-023-12120-7} {\bibfield  {journal}
  {\bibinfo  {journal} {Eur. Phys. J. C}\ }\textbf {\bibinfo {volume} {83}},\
  \bibinfo {pages} {947} (\bibinfo {year} {2023})},\ \Eprint
  {https://arxiv.org/abs/2306.17513} {arXiv:2306.17513 [hep-ph]} \BibitemShut
  {NoStop}%
\bibitem [{\citenamefont {Caucal}\ \emph
  {et~al.}(2026{\natexlab{a}})\citenamefont {Caucal}, \citenamefont {Kang},
  \citenamefont {Korcyl}, \citenamefont {Salazar}, \citenamefont {Schenke},
  \citenamefont {Stebel}, \citenamefont {Venugopalan},\ and\ \citenamefont
  {Zhao}}]{Caucal:2025zkl}%
  \BibitemOpen
  \bibfield  {author} {\bibinfo {author} {\bibfnamefont {P.}~\bibnamefont
  {Caucal}}, \bibinfo {author} {\bibfnamefont {Z.-B.}\ \bibnamefont {Kang}},
  \bibinfo {author} {\bibfnamefont {P.}~\bibnamefont {Korcyl}}, \bibinfo
  {author} {\bibfnamefont {F.}~\bibnamefont {Salazar}}, \bibinfo {author}
  {\bibfnamefont {B.}~\bibnamefont {Schenke}}, \bibinfo {author} {\bibfnamefont
  {T.}~\bibnamefont {Stebel}}, \bibinfo {author} {\bibfnamefont
  {R.}~\bibnamefont {Venugopalan}},\ and\ \bibinfo {author} {\bibfnamefont
  {W.}~\bibnamefont {Zhao}},\ }\href
  {https://doi.org/10.1016/j.physletb.2026.140599} {\bibfield  {journal}
  {\bibinfo  {journal} {Phys. Lett. B}\ }\textbf {\bibinfo {volume} {879}},\
  \bibinfo {pages} {140599} (\bibinfo {year} {2026}{\natexlab{a}})},\ \Eprint
  {https://arxiv.org/abs/2512.21466} {arXiv:2512.21466 [hep-ph]} \BibitemShut
  {NoStop}%
\bibitem [{\citenamefont {Bermudez~Martinez}\ \emph {et~al.}(2019)\citenamefont
  {Bermudez~Martinez} \emph {et~al.}}]{BermudezMartinez:2019anj}%
  \BibitemOpen
  \bibfield  {author} {\bibinfo {author} {\bibfnamefont {A.}~\bibnamefont
  {Bermudez~Martinez}} \emph {et~al.},\ }\href
  {https://doi.org/10.1103/PhysRevD.100.074027} {\bibfield  {journal} {\bibinfo
   {journal} {Phys. Rev. D}\ }\textbf {\bibinfo {volume} {100}},\ \bibinfo
  {pages} {074027} (\bibinfo {year} {2019})},\ \Eprint
  {https://arxiv.org/abs/1906.00919} {arXiv:1906.00919 [hep-ph]} \BibitemShut
  {NoStop}%
\bibitem [{\citenamefont {Blanco}\ \emph {et~al.}(2019)\citenamefont {Blanco},
  \citenamefont {van Hameren}, \citenamefont {Jung}, \citenamefont {Kusina},\
  and\ \citenamefont {Kutak}}]{Blanco:2019qbm}%
  \BibitemOpen
  \bibfield  {author} {\bibinfo {author} {\bibfnamefont {E.}~\bibnamefont
  {Blanco}}, \bibinfo {author} {\bibfnamefont {A.}~\bibnamefont {van Hameren}},
  \bibinfo {author} {\bibfnamefont {H.}~\bibnamefont {Jung}}, \bibinfo {author}
  {\bibfnamefont {A.}~\bibnamefont {Kusina}},\ and\ \bibinfo {author}
  {\bibfnamefont {K.}~\bibnamefont {Kutak}},\ }\href
  {https://doi.org/10.1103/PhysRevD.100.054023} {\bibfield  {journal} {\bibinfo
   {journal} {Phys. Rev. D}\ }\textbf {\bibinfo {volume} {100}},\ \bibinfo
  {pages} {054023} (\bibinfo {year} {2019})},\ \Eprint
  {https://arxiv.org/abs/1905.07331} {arXiv:1905.07331 [hep-ph]} \BibitemShut
  {NoStop}%
\bibitem [{\citenamefont {Beni{\'c}}\ \emph {et~al.}(2022)\citenamefont
  {Beni{\'c}}, \citenamefont {Garcia-Montero},\ and\ \citenamefont
  {Perkov}}]{Benic:2022ixp}%
  \BibitemOpen
  \bibfield  {author} {\bibinfo {author} {\bibfnamefont {S.}~\bibnamefont
  {Beni{\'c}}}, \bibinfo {author} {\bibfnamefont {O.}~\bibnamefont
  {Garcia-Montero}},\ and\ \bibinfo {author} {\bibfnamefont {A.}~\bibnamefont
  {Perkov}},\ }\href {https://doi.org/10.1103/PhysRevD.105.114052} {\bibfield
  {journal} {\bibinfo  {journal} {Phys. Rev. D}\ }\textbf {\bibinfo {volume}
  {105}},\ \bibinfo {pages} {114052} (\bibinfo {year} {2022})},\ \Eprint
  {https://arxiv.org/abs/2203.01685} {arXiv:2203.01685 [hep-ph]} \BibitemShut
  {NoStop}%
\bibitem [{\citenamefont {Marquet}\ \emph {et~al.}(2025)\citenamefont
  {Marquet}, \citenamefont {Shi},\ and\ \citenamefont
  {Xiao}}]{Marquet:2025jdr}%
  \BibitemOpen
  \bibfield  {author} {\bibinfo {author} {\bibfnamefont {C.}~\bibnamefont
  {Marquet}}, \bibinfo {author} {\bibfnamefont {Y.}~\bibnamefont {Shi}},\ and\
  \bibinfo {author} {\bibfnamefont {B.-W.}\ \bibnamefont {Xiao}},\ }\href@noop
  {} {\bibfield  {journal} {\bibinfo  {journal} {arXiv e-prints}\ } (\bibinfo
  {year} {2025})},\ \Eprint {https://arxiv.org/abs/2510.18949}
  {arXiv:2510.18949 [hep-ph]} \BibitemShut {NoStop}%
\bibitem [{\citenamefont {Gao}\ \emph {et~al.}(2026)\citenamefont {Gao},
  \citenamefont {Marquet}, \citenamefont {Shi},\ and\ \citenamefont
  {Xiao}}]{Gao:2026azd}%
  \BibitemOpen
  \bibfield  {author} {\bibinfo {author} {\bibfnamefont {Z.}~\bibnamefont
  {Gao}}, \bibinfo {author} {\bibfnamefont {C.}~\bibnamefont {Marquet}},
  \bibinfo {author} {\bibfnamefont {Y.}~\bibnamefont {Shi}},\ and\ \bibinfo
  {author} {\bibfnamefont {B.-W.}\ \bibnamefont {Xiao}},\ }\href@noop {}
  {\bibfield  {journal} {\bibinfo  {journal} {arXiv e-prints}\ } (\bibinfo
  {year} {2026})},\ \Eprint {https://arxiv.org/abs/2605.01527}
  {arXiv:2605.01527 [hep-ph]} \BibitemShut {NoStop}%
\bibitem [{\citenamefont {Mueller}\ \emph
  {et~al.}(2013{\natexlab{a}})\citenamefont {Mueller}, \citenamefont {Xiao},\
  and\ \citenamefont {Yuan}}]{Mueller:2012uf}%
  \BibitemOpen
  \bibfield  {author} {\bibinfo {author} {\bibfnamefont {A.~H.}\ \bibnamefont
  {Mueller}}, \bibinfo {author} {\bibfnamefont {B.-W.}\ \bibnamefont {Xiao}},\
  and\ \bibinfo {author} {\bibfnamefont {F.}~\bibnamefont {Yuan}},\ }\href
  {https://doi.org/10.1103/PhysRevLett.110.082301} {\bibfield  {journal}
  {\bibinfo  {journal} {Phys. Rev. Lett.}\ }\textbf {\bibinfo {volume} {110}},\
  \bibinfo {pages} {082301} (\bibinfo {year} {2013}{\natexlab{a}})},\ \Eprint
  {https://arxiv.org/abs/1210.5792} {arXiv:1210.5792 [hep-ph]} \BibitemShut
  {NoStop}%
\bibitem [{\citenamefont {Mueller}\ \emph
  {et~al.}(2013{\natexlab{b}})\citenamefont {Mueller}, \citenamefont {Xiao},\
  and\ \citenamefont {Yuan}}]{Mueller:2013wwa}%
  \BibitemOpen
  \bibfield  {author} {\bibinfo {author} {\bibfnamefont {A.~H.}\ \bibnamefont
  {Mueller}}, \bibinfo {author} {\bibfnamefont {B.-W.}\ \bibnamefont {Xiao}},\
  and\ \bibinfo {author} {\bibfnamefont {F.}~\bibnamefont {Yuan}},\ }\href
  {https://doi.org/10.1103/PhysRevD.88.114010} {\bibfield  {journal} {\bibinfo
  {journal} {Phys. Rev. D}\ }\textbf {\bibinfo {volume} {88}},\ \bibinfo
  {pages} {114010} (\bibinfo {year} {2013}{\natexlab{b}})},\ \Eprint
  {https://arxiv.org/abs/1308.2993} {arXiv:1308.2993 [hep-ph]} \BibitemShut
  {NoStop}%
\bibitem [{\citenamefont {Watanabe}\ and\ \citenamefont
  {Xiao}(2015)}]{Watanabe:2015yca}%
  \BibitemOpen
  \bibfield  {author} {\bibinfo {author} {\bibfnamefont {K.}~\bibnamefont
  {Watanabe}}\ and\ \bibinfo {author} {\bibfnamefont {B.-W.}\ \bibnamefont
  {Xiao}},\ }\href {https://doi.org/10.1103/PhysRevD.92.111502} {\bibfield
  {journal} {\bibinfo  {journal} {Phys. Rev. D}\ }\textbf {\bibinfo {volume}
  {92}},\ \bibinfo {pages} {111502} (\bibinfo {year} {2015})},\ \Eprint
  {https://arxiv.org/abs/1507.06564} {arXiv:1507.06564 [hep-ph]} \BibitemShut
  {NoStop}%
\bibitem [{\citenamefont {Xiao}\ \emph {et~al.}(2017)\citenamefont {Xiao},
  \citenamefont {Yuan},\ and\ \citenamefont {Zhou}}]{Xiao:2017yya}%
  \BibitemOpen
  \bibfield  {author} {\bibinfo {author} {\bibfnamefont {B.-W.}\ \bibnamefont
  {Xiao}}, \bibinfo {author} {\bibfnamefont {F.}~\bibnamefont {Yuan}},\ and\
  \bibinfo {author} {\bibfnamefont {J.}~\bibnamefont {Zhou}},\ }\href
  {https://doi.org/10.1016/j.nuclphysb.2017.05.012} {\bibfield  {journal}
  {\bibinfo  {journal} {Nucl. Phys. B}\ }\textbf {\bibinfo {volume} {921}},\
  \bibinfo {pages} {104} (\bibinfo {year} {2017})},\ \Eprint
  {https://arxiv.org/abs/1703.06163} {arXiv:1703.06163 [hep-ph]} \BibitemShut
  {NoStop}%
\bibitem [{\citenamefont {Zhou}(2019)}]{Zhou:2018lfq}%
  \BibitemOpen
  \bibfield  {author} {\bibinfo {author} {\bibfnamefont {J.}~\bibnamefont
  {Zhou}},\ }\href {https://doi.org/10.1103/PhysRevD.99.054026} {\bibfield
  {journal} {\bibinfo  {journal} {Phys. Rev. D}\ }\textbf {\bibinfo {volume}
  {99}},\ \bibinfo {pages} {054026} (\bibinfo {year} {2019})},\ \Eprint
  {https://arxiv.org/abs/1807.00506} {arXiv:1807.00506 [hep-ph]} \BibitemShut
  {NoStop}%
\bibitem [{\citenamefont {van Hameren}\ \emph {et~al.}(2021)\citenamefont {van
  Hameren}, \citenamefont {Kotko}, \citenamefont {Kutak},\ and\ \citenamefont
  {Sapeta}}]{vanHameren:2020rqt}%
  \BibitemOpen
  \bibfield  {author} {\bibinfo {author} {\bibfnamefont {A.}~\bibnamefont {van
  Hameren}}, \bibinfo {author} {\bibfnamefont {P.}~\bibnamefont {Kotko}},
  \bibinfo {author} {\bibfnamefont {K.}~\bibnamefont {Kutak}},\ and\ \bibinfo
  {author} {\bibfnamefont {S.}~\bibnamefont {Sapeta}},\ }\href
  {https://doi.org/10.1016/j.physletb.2021.136078} {\bibfield  {journal}
  {\bibinfo  {journal} {Phys. Lett. B}\ }\textbf {\bibinfo {volume} {814}},\
  \bibinfo {pages} {136078} (\bibinfo {year} {2021})},\ \Eprint
  {https://arxiv.org/abs/2010.13066} {arXiv:2010.13066 [hep-ph]} \BibitemShut
  {NoStop}%
\bibitem [{\citenamefont {Balitsky}(2021)}]{Balitsky:2020jzt}%
  \BibitemOpen
  \bibfield  {author} {\bibinfo {author} {\bibfnamefont {I.}~\bibnamefont
  {Balitsky}},\ }\href {https://doi.org/10.1007/JHEP05(2021)046} {\bibfield
  {journal} {\bibinfo  {journal} {JHEP}\ }\textbf {\bibinfo {volume}
  {2021}}\bibfield  {number} {\bibinfo  {number} { (05)},\ \bibinfo {pages}
  {046}},\ }\Eprint {https://arxiv.org/abs/2012.01588} {arXiv:2012.01588
  [hep-ph]} \BibitemShut {NoStop}%
\bibitem [{\citenamefont {Hentschinski}(2021)}]{Hentschinski:2021lsh}%
  \BibitemOpen
  \bibfield  {author} {\bibinfo {author} {\bibfnamefont {M.}~\bibnamefont
  {Hentschinski}},\ }\href {https://doi.org/10.1103/PhysRevD.104.054014}
  {\bibfield  {journal} {\bibinfo  {journal} {Phys. Rev. D}\ }\textbf {\bibinfo
  {volume} {104}},\ \bibinfo {pages} {054014} (\bibinfo {year} {2021})},\
  \Eprint {https://arxiv.org/abs/2107.06203} {arXiv:2107.06203 [hep-ph]}
  \BibitemShut {NoStop}%
\bibitem [{\citenamefont {Goda}\ \emph {et~al.}(2023)\citenamefont {Goda},
  \citenamefont {Kutak},\ and\ \citenamefont {Sapeta}}]{Goda:2022wsc}%
  \BibitemOpen
  \bibfield  {author} {\bibinfo {author} {\bibfnamefont {T.}~\bibnamefont
  {Goda}}, \bibinfo {author} {\bibfnamefont {K.}~\bibnamefont {Kutak}},\ and\
  \bibinfo {author} {\bibfnamefont {S.}~\bibnamefont {Sapeta}},\ }\href
  {https://doi.org/10.1016/j.nuclphysb.2023.116155} {\bibfield  {journal}
  {\bibinfo  {journal} {Nucl. Phys. B}\ }\textbf {\bibinfo {volume} {990}},\
  \bibinfo {pages} {116155} (\bibinfo {year} {2023})},\ \Eprint
  {https://arxiv.org/abs/2210.16084} {arXiv:2210.16084 [hep-ph]} \BibitemShut
  {NoStop}%
\bibitem [{\citenamefont {Mukherjee}\ \emph {et~al.}(2024)\citenamefont
  {Mukherjee}, \citenamefont {Skokov}, \citenamefont {Tarasov},\ and\
  \citenamefont {Tiwari}}]{Mukherjee:2023snp}%
  \BibitemOpen
  \bibfield  {author} {\bibinfo {author} {\bibfnamefont {S.}~\bibnamefont
  {Mukherjee}}, \bibinfo {author} {\bibfnamefont {V.~V.}\ \bibnamefont
  {Skokov}}, \bibinfo {author} {\bibfnamefont {A.}~\bibnamefont {Tarasov}},\
  and\ \bibinfo {author} {\bibfnamefont {S.}~\bibnamefont {Tiwari}},\ }\href
  {https://doi.org/10.1103/PhysRevD.109.034035} {\bibfield  {journal} {\bibinfo
   {journal} {Phys. Rev. D}\ }\textbf {\bibinfo {volume} {109}},\ \bibinfo
  {pages} {034035} (\bibinfo {year} {2024})},\ \Eprint
  {https://arxiv.org/abs/2311.16402} {arXiv:2311.16402 [hep-ph]} \BibitemShut
  {NoStop}%
\bibitem [{\citenamefont {Altinoluk}\ \emph {et~al.}(2024)\citenamefont
  {Altinoluk}, \citenamefont {Jalilian-Marian},\ and\ \citenamefont
  {Marquet}}]{Altinoluk:2024vgg}%
  \BibitemOpen
  \bibfield  {author} {\bibinfo {author} {\bibfnamefont {T.}~\bibnamefont
  {Altinoluk}}, \bibinfo {author} {\bibfnamefont {J.}~\bibnamefont
  {Jalilian-Marian}},\ and\ \bibinfo {author} {\bibfnamefont {C.}~\bibnamefont
  {Marquet}},\ }\href {https://doi.org/10.1103/PhysRevD.110.094056} {\bibfield
  {journal} {\bibinfo  {journal} {Phys. Rev. D}\ }\textbf {\bibinfo {volume}
  {110}},\ \bibinfo {pages} {094056} (\bibinfo {year} {2024})},\ \Eprint
  {https://arxiv.org/abs/2406.08277} {arXiv:2406.08277 [hep-ph]} \BibitemShut
  {NoStop}%
\bibitem [{\citenamefont {Duan}\ \emph {et~al.}(2025)\citenamefont {Duan},
  \citenamefont {Kovner},\ and\ \citenamefont {Lublinsky}}]{Duan:2024nlr}%
  \BibitemOpen
  \bibfield  {author} {\bibinfo {author} {\bibfnamefont {H.}~\bibnamefont
  {Duan}}, \bibinfo {author} {\bibfnamefont {A.}~\bibnamefont {Kovner}},\ and\
  \bibinfo {author} {\bibfnamefont {M.}~\bibnamefont {Lublinsky}},\ }\href
  {https://doi.org/10.1103/PhysRevD.111.054022} {\bibfield  {journal} {\bibinfo
   {journal} {Phys. Rev. D}\ }\textbf {\bibinfo {volume} {111}},\ \bibinfo
  {pages} {054022} (\bibinfo {year} {2025})},\ \Eprint
  {https://arxiv.org/abs/2407.15960} {arXiv:2407.15960 [hep-ph]} \BibitemShut
  {NoStop}%
\bibitem [{\citenamefont {van Hameren}\ and\ \citenamefont
  {Nefedov}(2025)}]{vanHameren:2025hyo}%
  \BibitemOpen
  \bibfield  {author} {\bibinfo {author} {\bibfnamefont {A.}~\bibnamefont {van
  Hameren}}\ and\ \bibinfo {author} {\bibfnamefont {M.}~\bibnamefont
  {Nefedov}},\ }\href {https://doi.org/10.1007/JHEP02(2025)160} {\bibfield
  {journal} {\bibinfo  {journal} {JHEP}\ }\textbf {\bibinfo {volume}
  {2025}}\bibfield  {number} {\bibinfo  {number} { (02)},\ \bibinfo {pages}
  {160}},\ }\Eprint {https://arxiv.org/abs/2501.02619} {arXiv:2501.02619
  [hep-ph]} \BibitemShut {NoStop}%
\bibitem [{\citenamefont {Balitsky}(2026)}]{Balitsky:2026lop}%
  \BibitemOpen
  \bibfield  {author} {\bibinfo {author} {\bibfnamefont {I.}~\bibnamefont
  {Balitsky}},\ }\href@noop {} {\bibfield  {journal} {\bibinfo  {journal}
  {arXiv e-print}\ } (\bibinfo {year} {2026})},\ \Eprint
  {https://arxiv.org/abs/2606.07429} {arXiv:2606.07429 [hep-ph]} \BibitemShut
  {NoStop}%
\bibitem [{\citenamefont {Caucal}\ \emph {et~al.}(2022)\citenamefont {Caucal},
  \citenamefont {Salazar}, \citenamefont {Schenke},\ and\ \citenamefont
  {Venugopalan}}]{Caucal:2022ulg}%
  \BibitemOpen
  \bibfield  {author} {\bibinfo {author} {\bibfnamefont {P.}~\bibnamefont
  {Caucal}}, \bibinfo {author} {\bibfnamefont {F.}~\bibnamefont {Salazar}},
  \bibinfo {author} {\bibfnamefont {B.}~\bibnamefont {Schenke}},\ and\ \bibinfo
  {author} {\bibfnamefont {R.}~\bibnamefont {Venugopalan}},\ }\href
  {https://doi.org/10.1007/JHEP11(2022)169} {\bibfield  {journal} {\bibinfo
  {journal} {JHEP}\ }\textbf {\bibinfo {volume} {2022}}\bibfield  {number}
  {\bibinfo  {number} { (11)},\ \bibinfo {pages} {169}},\ }\Eprint
  {https://arxiv.org/abs/2208.13872} {arXiv:2208.13872 [hep-ph]} \BibitemShut
  {NoStop}%
\bibitem [{\citenamefont {Taels}\ \emph {et~al.}(2022)\citenamefont {Taels},
  \citenamefont {Altinoluk}, \citenamefont {Beuf},\ and\ \citenamefont
  {Marquet}}]{Taels:2022tza}%
  \BibitemOpen
  \bibfield  {author} {\bibinfo {author} {\bibfnamefont {P.}~\bibnamefont
  {Taels}}, \bibinfo {author} {\bibfnamefont {T.}~\bibnamefont {Altinoluk}},
  \bibinfo {author} {\bibfnamefont {G.}~\bibnamefont {Beuf}},\ and\ \bibinfo
  {author} {\bibfnamefont {C.}~\bibnamefont {Marquet}},\ }\href
  {https://doi.org/10.1007/JHEP10(2022)184} {\bibfield  {journal} {\bibinfo
  {journal} {JHEP}\ }\textbf {\bibinfo {volume} {2022}}\bibfield  {number}
  {\bibinfo  {number} { (10)},\ \bibinfo {pages} {184}},\ }\Eprint
  {https://arxiv.org/abs/2204.11650} {arXiv:2204.11650 [hep-ph]} \BibitemShut
  {NoStop}%
\bibitem [{\citenamefont {Caucal}\ \emph {et~al.}(2023)\citenamefont {Caucal},
  \citenamefont {Salazar}, \citenamefont {Schenke}, \citenamefont {Stebel},\
  and\ \citenamefont {Venugopalan}}]{Caucal:2023nci}%
  \BibitemOpen
  \bibfield  {author} {\bibinfo {author} {\bibfnamefont {P.}~\bibnamefont
  {Caucal}}, \bibinfo {author} {\bibfnamefont {F.}~\bibnamefont {Salazar}},
  \bibinfo {author} {\bibfnamefont {B.}~\bibnamefont {Schenke}}, \bibinfo
  {author} {\bibfnamefont {T.}~\bibnamefont {Stebel}},\ and\ \bibinfo {author}
  {\bibfnamefont {R.}~\bibnamefont {Venugopalan}},\ }\href
  {https://doi.org/10.1007/JHEP08(2023)062} {\bibfield  {journal} {\bibinfo
  {journal} {JHEP}\ }\textbf {\bibinfo {volume} {2023}}\bibfield  {number}
  {\bibinfo  {number} { (08)},\ \bibinfo {pages} {062}},\ }\Eprint
  {https://arxiv.org/abs/2304.03304} {arXiv:2304.03304 [hep-ph]} \BibitemShut
  {NoStop}%
\bibitem [{\citenamefont {Caucal}\ \emph {et~al.}(2024)\citenamefont {Caucal},
  \citenamefont {Salazar}, \citenamefont {Schenke}, \citenamefont {Stebel},\
  and\ \citenamefont {Venugopalan}}]{Caucal:2023fsf}%
  \BibitemOpen
  \bibfield  {author} {\bibinfo {author} {\bibfnamefont {P.}~\bibnamefont
  {Caucal}}, \bibinfo {author} {\bibfnamefont {F.}~\bibnamefont {Salazar}},
  \bibinfo {author} {\bibfnamefont {B.}~\bibnamefont {Schenke}}, \bibinfo
  {author} {\bibfnamefont {T.}~\bibnamefont {Stebel}},\ and\ \bibinfo {author}
  {\bibfnamefont {R.}~\bibnamefont {Venugopalan}},\ }\href
  {https://doi.org/10.1103/PhysRevLett.132.081902} {\bibfield  {journal}
  {\bibinfo  {journal} {Phys. Rev. Lett.}\ }\textbf {\bibinfo {volume} {132}},\
  \bibinfo {pages} {081902} (\bibinfo {year} {2024})},\ \Eprint
  {https://arxiv.org/abs/2308.00022} {arXiv:2308.00022 [hep-ph]} \BibitemShut
  {NoStop}%
\bibitem [{\citenamefont {Caucal}\ and\ \citenamefont
  {Salazar}(2024)}]{Caucal:2024nsb}%
  \BibitemOpen
  \bibfield  {author} {\bibinfo {author} {\bibfnamefont {P.}~\bibnamefont
  {Caucal}}\ and\ \bibinfo {author} {\bibfnamefont {F.}~\bibnamefont
  {Salazar}},\ }\href {https://doi.org/10.1007/JHEP12(2024)130} {\bibfield
  {journal} {\bibinfo  {journal} {JHEP}\ }\textbf {\bibinfo {volume}
  {2024}}\bibfield  {number} {\bibinfo  {number} { (12)},\ \bibinfo {pages}
  {130}},\ }\Eprint {https://arxiv.org/abs/2405.19404} {arXiv:2405.19404
  [hep-ph]} \BibitemShut {NoStop}%
\bibitem [{\citenamefont {Bijl}\ \emph {et~al.}(2024)\citenamefont {Bijl},
  \citenamefont {Niedenzu},\ and\ \citenamefont {Waalewijn}}]{Bijl:2023dux}%
  \BibitemOpen
  \bibfield  {author} {\bibinfo {author} {\bibfnamefont {P.}~\bibnamefont
  {Bijl}}, \bibinfo {author} {\bibfnamefont {S.}~\bibnamefont {Niedenzu}},\
  and\ \bibinfo {author} {\bibfnamefont {W.~J.}\ \bibnamefont {Waalewijn}},\
  }\href {https://doi.org/10.1103/PhysRevD.109.014011} {\bibfield  {journal}
  {\bibinfo  {journal} {Phys. Rev. D}\ }\textbf {\bibinfo {volume} {109}},\
  \bibinfo {pages} {014011} (\bibinfo {year} {2024})},\ \Eprint
  {https://arxiv.org/abs/2307.02521} {arXiv:2307.02521 [hep-ph]} \BibitemShut
  {NoStop}%
\bibitem [{\citenamefont {Hautmann}\ \emph {et~al.}(2012)\citenamefont
  {Hautmann}, \citenamefont {Hentschinski},\ and\ \citenamefont
  {Jung}}]{Hautmann:2012sh}%
  \BibitemOpen
  \bibfield  {author} {\bibinfo {author} {\bibfnamefont {F.}~\bibnamefont
  {Hautmann}}, \bibinfo {author} {\bibfnamefont {M.}~\bibnamefont
  {Hentschinski}},\ and\ \bibinfo {author} {\bibfnamefont {H.}~\bibnamefont
  {Jung}},\ }\href {https://doi.org/10.1016/j.nuclphysb.2012.07.023} {\bibfield
   {journal} {\bibinfo  {journal} {Nucl. Phys. B}\ }\textbf {\bibinfo {volume}
  {865}},\ \bibinfo {pages} {54} (\bibinfo {year} {2012})},\ \Eprint
  {https://arxiv.org/abs/1205.1759} {arXiv:1205.1759 [hep-ph]} \BibitemShut
  {NoStop}%
\bibitem [{\citenamefont {Marquet}\ \emph {et~al.}(2020)\citenamefont
  {Marquet}, \citenamefont {Wei},\ and\ \citenamefont
  {Xiao}}]{Marquet:2019ltn}%
  \BibitemOpen
  \bibfield  {author} {\bibinfo {author} {\bibfnamefont {C.}~\bibnamefont
  {Marquet}}, \bibinfo {author} {\bibfnamefont {S.-Y.}\ \bibnamefont {Wei}},\
  and\ \bibinfo {author} {\bibfnamefont {B.-W.}\ \bibnamefont {Xiao}},\ }\href
  {https://doi.org/10.1016/j.physletb.2020.135253} {\bibfield  {journal}
  {\bibinfo  {journal} {Phys. Lett. B}\ }\textbf {\bibinfo {volume} {802}},\
  \bibinfo {pages} {135253} (\bibinfo {year} {2020})},\ \Eprint
  {https://arxiv.org/abs/1909.08572} {arXiv:1909.08572 [hep-ph]} \BibitemShut
  {NoStop}%
\bibitem [{\citenamefont {Stewart}\ \emph {et~al.}(2010)\citenamefont
  {Stewart}, \citenamefont {Tackmann},\ and\ \citenamefont
  {Waalewijn}}]{Stewart:2010tn}%
  \BibitemOpen
  \bibfield  {author} {\bibinfo {author} {\bibfnamefont {I.~W.}\ \bibnamefont
  {Stewart}}, \bibinfo {author} {\bibfnamefont {F.~J.}\ \bibnamefont
  {Tackmann}},\ and\ \bibinfo {author} {\bibfnamefont {W.~J.}\ \bibnamefont
  {Waalewijn}},\ }\href {https://doi.org/10.1103/PhysRevLett.105.092002}
  {\bibfield  {journal} {\bibinfo  {journal} {Phys. Rev. Lett.}\ }\textbf
  {\bibinfo {volume} {105}},\ \bibinfo {pages} {092002} (\bibinfo {year}
  {2010})},\ \Eprint {https://arxiv.org/abs/1004.2489} {arXiv:1004.2489
  [hep-ph]} \BibitemShut {NoStop}%
\bibitem [{\citenamefont {Kang}\ \emph {et~al.}(2012)\citenamefont {Kang},
  \citenamefont {Mantry},\ and\ \citenamefont {Qiu}}]{Kang:2012zr}%
  \BibitemOpen
  \bibfield  {author} {\bibinfo {author} {\bibfnamefont {Z.-B.}\ \bibnamefont
  {Kang}}, \bibinfo {author} {\bibfnamefont {S.}~\bibnamefont {Mantry}},\ and\
  \bibinfo {author} {\bibfnamefont {J.-W.}\ \bibnamefont {Qiu}},\ }\href
  {https://doi.org/10.1103/PhysRevD.86.114011} {\bibfield  {journal} {\bibinfo
  {journal} {Phys. Rev. D}\ }\textbf {\bibinfo {volume} {86}},\ \bibinfo
  {pages} {114011} (\bibinfo {year} {2012})},\ \Eprint
  {https://arxiv.org/abs/1204.5469} {arXiv:1204.5469 [hep-ph]} \BibitemShut
  {NoStop}%
\bibitem [{\citenamefont {Kang}\ \emph {et~al.}(2013)\citenamefont {Kang},
  \citenamefont {Liu}, \citenamefont {Mantry},\ and\ \citenamefont
  {Qiu}}]{Kang:2013wca}%
  \BibitemOpen
  \bibfield  {author} {\bibinfo {author} {\bibfnamefont {Z.-B.}\ \bibnamefont
  {Kang}}, \bibinfo {author} {\bibfnamefont {X.}~\bibnamefont {Liu}}, \bibinfo
  {author} {\bibfnamefont {S.}~\bibnamefont {Mantry}},\ and\ \bibinfo {author}
  {\bibfnamefont {J.-W.}\ \bibnamefont {Qiu}},\ }\href
  {https://doi.org/10.1103/PhysRevD.88.074020} {\bibfield  {journal} {\bibinfo
  {journal} {Phys. Rev. D}\ }\textbf {\bibinfo {volume} {88}},\ \bibinfo
  {pages} {074020} (\bibinfo {year} {2013})},\ \Eprint
  {https://arxiv.org/abs/1303.3063} {arXiv:1303.3063 [hep-ph]} \BibitemShut
  {NoStop}%
\bibitem [{\citenamefont {Procura}\ \emph {et~al.}(2015)\citenamefont
  {Procura}, \citenamefont {Waalewijn},\ and\ \citenamefont
  {Zeune}}]{Procura:2014cba}%
  \BibitemOpen
  \bibfield  {author} {\bibinfo {author} {\bibfnamefont {M.}~\bibnamefont
  {Procura}}, \bibinfo {author} {\bibfnamefont {W.~J.}\ \bibnamefont
  {Waalewijn}},\ and\ \bibinfo {author} {\bibfnamefont {L.}~\bibnamefont
  {Zeune}},\ }\href {https://doi.org/10.1007/JHEP02(2015)117} {\bibfield
  {journal} {\bibinfo  {journal} {JHEP}\ }\textbf {\bibinfo {volume}
  {2015}}\bibfield  {number} {\bibinfo  {number} { (02)},\ \bibinfo {pages}
  {117}},\ }\Eprint {https://arxiv.org/abs/1410.6483} {arXiv:1410.6483
  [hep-ph]} \BibitemShut {NoStop}%
\bibitem [{\citenamefont {Lustermans}\ \emph {et~al.}(2019)\citenamefont
  {Lustermans}, \citenamefont {Michel}, \citenamefont {Tackmann},\ and\
  \citenamefont {Waalewijn}}]{Lustermans:2019plv}%
  \BibitemOpen
  \bibfield  {author} {\bibinfo {author} {\bibfnamefont {G.}~\bibnamefont
  {Lustermans}}, \bibinfo {author} {\bibfnamefont {J.~K.~L.}\ \bibnamefont
  {Michel}}, \bibinfo {author} {\bibfnamefont {F.~J.}\ \bibnamefont
  {Tackmann}},\ and\ \bibinfo {author} {\bibfnamefont {W.~J.}\ \bibnamefont
  {Waalewijn}},\ }\href {https://doi.org/10.1007/JHEP03(2019)124} {\bibfield
  {journal} {\bibinfo  {journal} {JHEP}\ }\textbf {\bibinfo {volume}
  {2019}}\bibfield  {number} {\bibinfo  {number} { (03)},\ \bibinfo {pages}
  {124}},\ }\Eprint {https://arxiv.org/abs/1901.03331} {arXiv:1901.03331
  [hep-ph]} \BibitemShut {NoStop}%
\bibitem [{\citenamefont {Monni}\ \emph {et~al.}(2020)\citenamefont {Monni},
  \citenamefont {Rottoli},\ and\ \citenamefont {Torrielli}}]{Monni:2019yyr}%
  \BibitemOpen
  \bibfield  {author} {\bibinfo {author} {\bibfnamefont {P.~F.}\ \bibnamefont
  {Monni}}, \bibinfo {author} {\bibfnamefont {L.}~\bibnamefont {Rottoli}},\
  and\ \bibinfo {author} {\bibfnamefont {P.}~\bibnamefont {Torrielli}},\ }\href
  {https://doi.org/10.1103/PhysRevLett.124.252001} {\bibfield  {journal}
  {\bibinfo  {journal} {Phys. Rev. Lett.}\ }\textbf {\bibinfo {volume} {124}},\
  \bibinfo {pages} {252001} (\bibinfo {year} {2020})},\ \Eprint
  {https://arxiv.org/abs/1909.04704} {arXiv:1909.04704 [hep-ph]} \BibitemShut
  {NoStop}%
\bibitem [{\citenamefont {Makris}\ \emph {et~al.}(2021)\citenamefont {Makris},
  \citenamefont {Ringer},\ and\ \citenamefont {Waalewijn}}]{Makris:2020ltr}%
  \BibitemOpen
  \bibfield  {author} {\bibinfo {author} {\bibfnamefont {Y.}~\bibnamefont
  {Makris}}, \bibinfo {author} {\bibfnamefont {F.}~\bibnamefont {Ringer}},\
  and\ \bibinfo {author} {\bibfnamefont {W.~J.}\ \bibnamefont {Waalewijn}},\
  }\href {https://doi.org/10.1007/JHEP02(2021)070} {\bibfield  {journal}
  {\bibinfo  {journal} {JHEP}\ }\textbf {\bibinfo {volume} {2021}}\bibfield
  {number} {\bibinfo  {number} { (02)},\ \bibinfo {pages} {070}},\ }\Eprint
  {https://arxiv.org/abs/2009.11871} {arXiv:2009.11871 [hep-ph]} \BibitemShut
  {NoStop}%
\bibitem [{\citenamefont {Boglione}\ and\ \citenamefont
  {Simonelli}(2021)}]{Boglione:2020auc}%
  \BibitemOpen
  \bibfield  {author} {\bibinfo {author} {\bibfnamefont {M.}~\bibnamefont
  {Boglione}}\ and\ \bibinfo {author} {\bibfnamefont {A.}~\bibnamefont
  {Simonelli}},\ }\href {https://doi.org/10.1007/JHEP02(2021)076} {\bibfield
  {journal} {\bibinfo  {journal} {JHEP}\ }\textbf {\bibinfo {volume}
  {2021}}\bibfield  {number} {\bibinfo  {number} { (02)},\ \bibinfo {pages}
  {076}},\ }\Eprint {https://arxiv.org/abs/2011.07366} {arXiv:2011.07366
  [hep-ph]} \BibitemShut {NoStop}%
\bibitem [{\citenamefont {Boglione}\ and\ \citenamefont
  {Simonelli}(2022)}]{Boglione:2021wov}%
  \BibitemOpen
  \bibfield  {author} {\bibinfo {author} {\bibfnamefont {M.}~\bibnamefont
  {Boglione}}\ and\ \bibinfo {author} {\bibfnamefont {A.}~\bibnamefont
  {Simonelli}},\ }\href {https://doi.org/10.1007/JHEP02(2022)013} {\bibfield
  {journal} {\bibinfo  {journal} {JHEP}\ }\textbf {\bibinfo {volume}
  {2022}}\bibfield  {number} {\bibinfo  {number} { (02)},\ \bibinfo {pages}
  {013}},\ }\Eprint {https://arxiv.org/abs/2109.11497} {arXiv:2109.11497
  [hep-ph]} \BibitemShut {NoStop}%
\bibitem [{\citenamefont {Boglione}\ and\ \citenamefont
  {Simonelli}(2023)}]{Boglione:2023duo}%
  \BibitemOpen
  \bibfield  {author} {\bibinfo {author} {\bibfnamefont {M.}~\bibnamefont
  {Boglione}}\ and\ \bibinfo {author} {\bibfnamefont {A.}~\bibnamefont
  {Simonelli}},\ }\href {https://doi.org/10.1007/JHEP09(2023)006} {\bibfield
  {journal} {\bibinfo  {journal} {JHEP}\ }\textbf {\bibinfo {volume}
  {2023}}\bibfield  {number} {\bibinfo  {number} { (09)},\ \bibinfo {pages}
  {006}},\ }\Eprint {https://arxiv.org/abs/2306.02937} {arXiv:2306.02937
  [hep-ph]} \BibitemShut {NoStop}%
\bibitem [{\citenamefont {Fang}\ \emph {et~al.}(2026)\citenamefont {Fang},
  \citenamefont {Lin}, \citenamefont {Shao},\ and\ \citenamefont
  {Zhou}}]{Fang:2025dee}%
  \BibitemOpen
  \bibfield  {author} {\bibinfo {author} {\bibfnamefont {S.}~\bibnamefont
  {Fang}}, \bibinfo {author} {\bibfnamefont {S.}~\bibnamefont {Lin}}, \bibinfo
  {author} {\bibfnamefont {D.~Y.}\ \bibnamefont {Shao}},\ and\ \bibinfo
  {author} {\bibfnamefont {J.}~\bibnamefont {Zhou}},\ }\href
  {https://doi.org/10.1103/rvgc-sgv7} {\bibfield  {journal} {\bibinfo
  {journal} {Phys. Rev. Lett.}\ }\textbf {\bibinfo {volume} {136}},\ \bibinfo
  {pages} {021901} (\bibinfo {year} {2026})},\ \Eprint
  {https://arxiv.org/abs/2506.15962} {arXiv:2506.15962 [hep-ph]} \BibitemShut
  {NoStop}%
\bibitem [{\citenamefont {Michel}\ \emph {et~al.}(2019)\citenamefont {Michel},
  \citenamefont {Pietrulewicz},\ and\ \citenamefont
  {Tackmann}}]{Michel:2018hui}%
  \BibitemOpen
  \bibfield  {author} {\bibinfo {author} {\bibfnamefont {J.~K.~L.}\
  \bibnamefont {Michel}}, \bibinfo {author} {\bibfnamefont {P.}~\bibnamefont
  {Pietrulewicz}},\ and\ \bibinfo {author} {\bibfnamefont {F.~J.}\ \bibnamefont
  {Tackmann}},\ }\href {https://doi.org/10.1007/JHEP04(2019)142} {\bibfield
  {journal} {\bibinfo  {journal} {JHEP}\ }\textbf {\bibinfo {volume}
  {2019}}\bibfield  {number} {\bibinfo  {number} { (04)},\ \bibinfo {pages}
  {142}},\ }\Eprint {https://arxiv.org/abs/1810.12911} {arXiv:1810.12911
  [hep-ph]} \BibitemShut {NoStop}%
\bibitem [{\citenamefont {Bauer}\ \emph {et~al.}(2000)\citenamefont {Bauer},
  \citenamefont {Fleming},\ and\ \citenamefont {Luke}}]{Bauer:2000ew}%
  \BibitemOpen
  \bibfield  {author} {\bibinfo {author} {\bibfnamefont {C.~W.}\ \bibnamefont
  {Bauer}}, \bibinfo {author} {\bibfnamefont {S.}~\bibnamefont {Fleming}},\
  and\ \bibinfo {author} {\bibfnamefont {M.~E.}\ \bibnamefont {Luke}},\ }\href
  {https://doi.org/10.1103/PhysRevD.63.014006} {\bibfield  {journal} {\bibinfo
  {journal} {Phys. Rev.}\ }\textbf {\bibinfo {volume} {D63}},\ \bibinfo {pages}
  {014006} (\bibinfo {year} {2000})},\ \Eprint
  {https://arxiv.org/abs/hep-ph/0005275} {arXiv:hep-ph/0005275} \BibitemShut
  {NoStop}%
\bibitem [{\citenamefont {Bauer}\ \emph {et~al.}(2001)\citenamefont {Bauer},
  \citenamefont {Fleming}, \citenamefont {Pirjol},\ and\ \citenamefont
  {Stewart}}]{Bauer:2000yr}%
  \BibitemOpen
  \bibfield  {author} {\bibinfo {author} {\bibfnamefont {C.~W.}\ \bibnamefont
  {Bauer}}, \bibinfo {author} {\bibfnamefont {S.}~\bibnamefont {Fleming}},
  \bibinfo {author} {\bibfnamefont {D.}~\bibnamefont {Pirjol}},\ and\ \bibinfo
  {author} {\bibfnamefont {I.~W.}\ \bibnamefont {Stewart}},\ }\href
  {https://doi.org/10.1103/PhysRevD.63.114020} {\bibfield  {journal} {\bibinfo
  {journal} {Phys. Rev.}\ }\textbf {\bibinfo {volume} {D63}},\ \bibinfo {pages}
  {114020} (\bibinfo {year} {2001})},\ \Eprint
  {https://arxiv.org/abs/hep-ph/0011336} {arXiv:hep-ph/0011336 [hep-ph]}
  \BibitemShut {NoStop}%
\bibitem [{\citenamefont {Bauer}\ and\ \citenamefont
  {Stewart}(2001)}]{Bauer:2001ct}%
  \BibitemOpen
  \bibfield  {author} {\bibinfo {author} {\bibfnamefont {C.~W.}\ \bibnamefont
  {Bauer}}\ and\ \bibinfo {author} {\bibfnamefont {I.~W.}\ \bibnamefont
  {Stewart}},\ }\href {https://doi.org/10.1016/S0370-2693(01)00902-9}
  {\bibfield  {journal} {\bibinfo  {journal} {Phys. Lett.}\ }\textbf {\bibinfo
  {volume} {B516}},\ \bibinfo {pages} {134} (\bibinfo {year} {2001})},\ \Eprint
  {https://arxiv.org/abs/hep-ph/0107001} {arXiv:hep-ph/0107001 [hep-ph]}
  \BibitemShut {NoStop}%
\bibitem [{\citenamefont {Bauer}\ \emph {et~al.}(2002)\citenamefont {Bauer},
  \citenamefont {Pirjol},\ and\ \citenamefont {Stewart}}]{Bauer:2001yt}%
  \BibitemOpen
  \bibfield  {author} {\bibinfo {author} {\bibfnamefont {C.~W.}\ \bibnamefont
  {Bauer}}, \bibinfo {author} {\bibfnamefont {D.}~\bibnamefont {Pirjol}},\ and\
  \bibinfo {author} {\bibfnamefont {I.~W.}\ \bibnamefont {Stewart}},\ }\href
  {https://doi.org/10.1103/PhysRevD.65.054022} {\bibfield  {journal} {\bibinfo
  {journal} {Phys. Rev.}\ }\textbf {\bibinfo {volume} {D65}},\ \bibinfo {pages}
  {054022} (\bibinfo {year} {2002})},\ \Eprint
  {https://arxiv.org/abs/hep-ph/0109045} {arXiv:hep-ph/0109045 [hep-ph]}
  \BibitemShut {NoStop}%
\bibitem [{\citenamefont {Gelis}\ and\ \citenamefont
  {Jalilian-Marian}(2003)}]{Gelis:2002nn}%
  \BibitemOpen
  \bibfield  {author} {\bibinfo {author} {\bibfnamefont {F.}~\bibnamefont
  {Gelis}}\ and\ \bibinfo {author} {\bibfnamefont {J.}~\bibnamefont
  {Jalilian-Marian}},\ }\href {https://doi.org/10.1103/PhysRevD.67.074019}
  {\bibfield  {journal} {\bibinfo  {journal} {Phys. Rev. D}\ }\textbf {\bibinfo
  {volume} {67}},\ \bibinfo {pages} {074019} (\bibinfo {year} {2003})},\
  \Eprint {https://arxiv.org/abs/hep-ph/0211363} {arXiv:hep-ph/0211363}
  \BibitemShut {NoStop}%
\bibitem [{\citenamefont {Altinoluk}\ and\ \citenamefont
  {Kovner}(2011)}]{Altinoluk:2011qy}%
  \BibitemOpen
  \bibfield  {author} {\bibinfo {author} {\bibfnamefont {T.}~\bibnamefont
  {Altinoluk}}\ and\ \bibinfo {author} {\bibfnamefont {A.}~\bibnamefont
  {Kovner}},\ }\href {https://doi.org/10.1103/PhysRevD.83.105004} {\bibfield
  {journal} {\bibinfo  {journal} {Phys. Rev. D}\ }\textbf {\bibinfo {volume}
  {83}},\ \bibinfo {pages} {105004} (\bibinfo {year} {2011})},\ \Eprint
  {https://arxiv.org/abs/1102.5327} {arXiv:1102.5327 [hep-ph]} \BibitemShut
  {NoStop}%
\bibitem [{\citenamefont {Chirilli}\ \emph {et~al.}(2012)\citenamefont
  {Chirilli}, \citenamefont {Xiao},\ and\ \citenamefont
  {Yuan}}]{Chirilli:2011km}%
  \BibitemOpen
  \bibfield  {author} {\bibinfo {author} {\bibfnamefont {G.~A.}\ \bibnamefont
  {Chirilli}}, \bibinfo {author} {\bibfnamefont {B.-W.}\ \bibnamefont {Xiao}},\
  and\ \bibinfo {author} {\bibfnamefont {F.}~\bibnamefont {Yuan}},\ }\href
  {https://doi.org/10.1103/PhysRevLett.108.122301} {\bibfield  {journal}
  {\bibinfo  {journal} {Phys. Rev. Lett.}\ }\textbf {\bibinfo {volume} {108}},\
  \bibinfo {pages} {122301} (\bibinfo {year} {2012})},\ \Eprint
  {https://arxiv.org/abs/1112.1061} {arXiv:1112.1061 [hep-ph]} \BibitemShut
  {NoStop}%
\bibitem [{\citenamefont {Kopeliovich}\ \emph {et~al.}(2001)\citenamefont
  {Kopeliovich}, \citenamefont {Raufeisen},\ and\ \citenamefont
  {Tarasov}}]{Kopeliovich:2000fb}%
  \BibitemOpen
  \bibfield  {author} {\bibinfo {author} {\bibfnamefont {B.~Z.}\ \bibnamefont
  {Kopeliovich}}, \bibinfo {author} {\bibfnamefont {J.}~\bibnamefont
  {Raufeisen}},\ and\ \bibinfo {author} {\bibfnamefont {A.~V.}\ \bibnamefont
  {Tarasov}},\ }\href {https://doi.org/10.1016/S0370-2693(01)00214-3}
  {\bibfield  {journal} {\bibinfo  {journal} {Phys. Lett. B}\ }\textbf
  {\bibinfo {volume} {503}},\ \bibinfo {pages} {91} (\bibinfo {year} {2001})},\
  \Eprint {https://arxiv.org/abs/hep-ph/0012035} {arXiv:hep-ph/0012035}
  \BibitemShut {NoStop}%
\bibitem [{\citenamefont {Gelis}\ and\ \citenamefont
  {Jalilian-Marian}(2002)}]{Gelis:2002fw}%
  \BibitemOpen
  \bibfield  {author} {\bibinfo {author} {\bibfnamefont {F.}~\bibnamefont
  {Gelis}}\ and\ \bibinfo {author} {\bibfnamefont {J.}~\bibnamefont
  {Jalilian-Marian}},\ }\href {https://doi.org/10.1103/PhysRevD.66.094014}
  {\bibfield  {journal} {\bibinfo  {journal} {Phys. Rev. D}\ }\textbf {\bibinfo
  {volume} {66}},\ \bibinfo {pages} {094014} (\bibinfo {year} {2002})},\
  \Eprint {https://arxiv.org/abs/hep-ph/0208141} {arXiv:hep-ph/0208141}
  \BibitemShut {NoStop}%
\bibitem [{\citenamefont {Baier}\ \emph {et~al.}(2004)\citenamefont {Baier},
  \citenamefont {Mueller},\ and\ \citenamefont {Schiff}}]{Baier:2004tj}%
  \BibitemOpen
  \bibfield  {author} {\bibinfo {author} {\bibfnamefont {R.}~\bibnamefont
  {Baier}}, \bibinfo {author} {\bibfnamefont {A.~H.}\ \bibnamefont {Mueller}},\
  and\ \bibinfo {author} {\bibfnamefont {D.}~\bibnamefont {Schiff}},\ }\href
  {https://doi.org/10.1016/j.nuclphysa.2004.06.020} {\bibfield  {journal}
  {\bibinfo  {journal} {Nucl. Phys. A}\ }\textbf {\bibinfo {volume} {741}},\
  \bibinfo {pages} {358} (\bibinfo {year} {2004})},\ \Eprint
  {https://arxiv.org/abs/hep-ph/0403201} {arXiv:hep-ph/0403201} \BibitemShut
  {NoStop}%
\bibitem [{\citenamefont {Gelis}\ and\ \citenamefont
  {Jalilian-Marian}(2007)}]{Gelis:2006hy}%
  \BibitemOpen
  \bibfield  {author} {\bibinfo {author} {\bibfnamefont {F.}~\bibnamefont
  {Gelis}}\ and\ \bibinfo {author} {\bibfnamefont {J.}~\bibnamefont
  {Jalilian-Marian}},\ }\href {https://doi.org/10.1103/PhysRevD.76.074015}
  {\bibfield  {journal} {\bibinfo  {journal} {Phys. Rev. D}\ }\textbf {\bibinfo
  {volume} {76}},\ \bibinfo {pages} {074015} (\bibinfo {year} {2007})},\
  \Eprint {https://arxiv.org/abs/hep-ph/0609066} {arXiv:hep-ph/0609066}
  \BibitemShut {NoStop}%
\bibitem [{\citenamefont {Golec-Biernat}\ \emph {et~al.}(2010)\citenamefont
  {Golec-Biernat}, \citenamefont {Lewandowska},\ and\ \citenamefont
  {Stasto}}]{GolecBiernat:2010de}%
  \BibitemOpen
  \bibfield  {author} {\bibinfo {author} {\bibfnamefont {K.}~\bibnamefont
  {Golec-Biernat}}, \bibinfo {author} {\bibfnamefont {E.}~\bibnamefont
  {Lewandowska}},\ and\ \bibinfo {author} {\bibfnamefont {A.~M.}\ \bibnamefont
  {Stasto}},\ }\href {https://doi.org/10.1103/PhysRevD.82.094010} {\bibfield
  {journal} {\bibinfo  {journal} {Phys. Rev. D}\ }\textbf {\bibinfo {volume}
  {82}},\ \bibinfo {pages} {094010} (\bibinfo {year} {2010})},\ \Eprint
  {https://arxiv.org/abs/1008.2652} {arXiv:1008.2652 [hep-ph]} \BibitemShut
  {NoStop}%
\bibitem [{\citenamefont {Duclou{\'e}}(2017)}]{Ducloue:2017zfd}%
  \BibitemOpen
  \bibfield  {author} {\bibinfo {author} {\bibfnamefont {B.}~\bibnamefont
  {Duclou{\'e}}},\ }\href {https://doi.org/10.1103/PhysRevD.96.094014}
  {\bibfield  {journal} {\bibinfo  {journal} {Phys. Rev. D}\ }\textbf {\bibinfo
  {volume} {96}},\ \bibinfo {pages} {094014} (\bibinfo {year} {2017})},\
  \Eprint {https://arxiv.org/abs/1701.08730} {arXiv:1701.08730 [hep-ph]}
  \BibitemShut {NoStop}%
\bibitem [{\citenamefont {Balitsky}\ and\ \citenamefont
  {Tarasov}(2016)}]{Balitsky:2016dgz}%
  \BibitemOpen
  \bibfield  {author} {\bibinfo {author} {\bibfnamefont {I.}~\bibnamefont
  {Balitsky}}\ and\ \bibinfo {author} {\bibfnamefont {A.}~\bibnamefont
  {Tarasov}},\ }\href {https://doi.org/10.1007/JHEP06(2016)164} {\bibfield
  {journal} {\bibinfo  {journal} {JHEP}\ }\textbf {\bibinfo {volume}
  {2016}}\bibfield  {number} {\bibinfo  {number} { (06)},\ \bibinfo {pages}
  {164}},\ }\Eprint {https://arxiv.org/abs/1603.06548} {arXiv:1603.06548
  [hep-ph]} \BibitemShut {NoStop}%
\bibitem [{\citenamefont {Altinoluk}\ and\ \citenamefont
  {Boussarie}(2019)}]{Altinoluk:2019wyu}%
  \BibitemOpen
  \bibfield  {author} {\bibinfo {author} {\bibfnamefont {T.}~\bibnamefont
  {Altinoluk}}\ and\ \bibinfo {author} {\bibfnamefont {R.}~\bibnamefont
  {Boussarie}},\ }\href {https://doi.org/10.1007/JHEP10(2019)208} {\bibfield
  {journal} {\bibinfo  {journal} {JHEP}\ }\textbf {\bibinfo {volume}
  {2019}}\bibfield  {number} {\bibinfo  {number} { (10)},\ \bibinfo {pages}
  {208}},\ }\Eprint {https://arxiv.org/abs/1902.07930} {arXiv:1902.07930
  [hep-ph]} \BibitemShut {NoStop}%
\bibitem [{\citenamefont {Caucal}\ \emph
  {et~al.}(2026{\natexlab{b}})\citenamefont {Caucal}, \citenamefont {Morales},
  \citenamefont {Iancu}, \citenamefont {Salazar},\ and\ \citenamefont
  {Yuan}}]{Caucal:2025xxh}%
  \BibitemOpen
  \bibfield  {author} {\bibinfo {author} {\bibfnamefont {P.}~\bibnamefont
  {Caucal}}, \bibinfo {author} {\bibfnamefont {M.~G.}\ \bibnamefont {Morales}},
  \bibinfo {author} {\bibfnamefont {E.}~\bibnamefont {Iancu}}, \bibinfo
  {author} {\bibfnamefont {F.}~\bibnamefont {Salazar}},\ and\ \bibinfo {author}
  {\bibfnamefont {F.}~\bibnamefont {Yuan}},\ }\href
  {https://doi.org/10.1016/j.physletb.2026.140271} {\bibfield  {journal}
  {\bibinfo  {journal} {Phys. Lett. B}\ }\textbf {\bibinfo {volume} {874}},\
  \bibinfo {pages} {140271} (\bibinfo {year} {2026}{\natexlab{b}})},\ \Eprint
  {https://arxiv.org/abs/2503.16162} {arXiv:2503.16162 [hep-ph]} \BibitemShut
  {NoStop}%
\bibitem [{\citenamefont {Marquet}\ \emph {et~al.}(2009)\citenamefont
  {Marquet}, \citenamefont {Xiao},\ and\ \citenamefont
  {Yuan}}]{Marquet:2009ca}%
  \BibitemOpen
  \bibfield  {author} {\bibinfo {author} {\bibfnamefont {C.}~\bibnamefont
  {Marquet}}, \bibinfo {author} {\bibfnamefont {B.-W.}\ \bibnamefont {Xiao}},\
  and\ \bibinfo {author} {\bibfnamefont {F.}~\bibnamefont {Yuan}},\ }\href
  {https://doi.org/10.1016/j.physletb.2009.10.099} {\bibfield  {journal}
  {\bibinfo  {journal} {Phys. Lett. B}\ }\textbf {\bibinfo {volume} {682}},\
  \bibinfo {pages} {207} (\bibinfo {year} {2009})},\ \Eprint
  {https://arxiv.org/abs/0906.1454} {arXiv:0906.1454 [hep-ph]} \BibitemShut
  {NoStop}%
\bibitem [{\citenamefont {Xiao}\ and\ \citenamefont
  {Yuan}(2010)}]{Xiao:2010sa}%
  \BibitemOpen
  \bibfield  {author} {\bibinfo {author} {\bibfnamefont {B.-W.}\ \bibnamefont
  {Xiao}}\ and\ \bibinfo {author} {\bibfnamefont {F.}~\bibnamefont {Yuan}},\
  }\href {https://doi.org/10.1103/PhysRevD.82.114009} {\bibfield  {journal}
  {\bibinfo  {journal} {Phys. Rev. D}\ }\textbf {\bibinfo {volume} {82}},\
  \bibinfo {pages} {114009} (\bibinfo {year} {2010})},\ \Eprint
  {https://arxiv.org/abs/1008.4432} {arXiv:1008.4432 [hep-ph]} \BibitemShut
  {NoStop}%
\bibitem [{\citenamefont {Dominguez}\ \emph {et~al.}(2011)\citenamefont
  {Dominguez}, \citenamefont {Marquet}, \citenamefont {Xiao},\ and\
  \citenamefont {Yuan}}]{Dominguez:2011wm}%
  \BibitemOpen
  \bibfield  {author} {\bibinfo {author} {\bibfnamefont {F.}~\bibnamefont
  {Dominguez}}, \bibinfo {author} {\bibfnamefont {C.}~\bibnamefont {Marquet}},
  \bibinfo {author} {\bibfnamefont {B.-W.}\ \bibnamefont {Xiao}},\ and\
  \bibinfo {author} {\bibfnamefont {F.}~\bibnamefont {Yuan}},\ }\href
  {https://doi.org/10.1103/PhysRevD.83.105005} {\bibfield  {journal} {\bibinfo
  {journal} {Phys.Rev.}\ }\textbf {\bibinfo {volume} {D83}},\ \bibinfo {pages}
  {105005} (\bibinfo {year} {2011})},\ \Eprint
  {https://arxiv.org/abs/1101.0715} {arXiv:1101.0715 [hep-ph]} \BibitemShut
  {NoStop}%
\bibitem [{\citenamefont {Caucal}\ \emph {et~al.}(2025)\citenamefont {Caucal},
  \citenamefont {Iancu}, \citenamefont {Mueller},\ and\ \citenamefont
  {Yuan}}]{Caucal:2024vbv}%
  \BibitemOpen
  \bibfield  {author} {\bibinfo {author} {\bibfnamefont {P.}~\bibnamefont
  {Caucal}}, \bibinfo {author} {\bibfnamefont {E.}~\bibnamefont {Iancu}},
  \bibinfo {author} {\bibfnamefont {A.~H.}\ \bibnamefont {Mueller}},\ and\
  \bibinfo {author} {\bibfnamefont {F.}~\bibnamefont {Yuan}},\ }\href
  {https://doi.org/10.1103/PhysRevLett.134.061903} {\bibfield  {journal}
  {\bibinfo  {journal} {Phys. Rev. Lett.}\ }\textbf {\bibinfo {volume} {134}},\
  \bibinfo {pages} {061903} (\bibinfo {year} {2025})},\ \Eprint
  {https://arxiv.org/abs/2408.03129} {arXiv:2408.03129 [hep-ph]} \BibitemShut
  {NoStop}%
\bibitem [{\citenamefont {Balitsky}(1996)}]{Balitsky:1995ub}%
  \BibitemOpen
  \bibfield  {author} {\bibinfo {author} {\bibfnamefont {I.}~\bibnamefont
  {Balitsky}},\ }\href {https://doi.org/10.1016/0550-3213(95)00638-9}
  {\bibfield  {journal} {\bibinfo  {journal} {Nucl. Phys. B}\ }\textbf
  {\bibinfo {volume} {463}},\ \bibinfo {pages} {99} (\bibinfo {year} {1996})},\
  \Eprint {https://arxiv.org/abs/hep-ph/9509348} {arXiv:hep-ph/9509348}
  \BibitemShut {NoStop}%
\bibitem [{\citenamefont {Kovchegov}(1999)}]{Kovchegov:1999yj}%
  \BibitemOpen
  \bibfield  {author} {\bibinfo {author} {\bibfnamefont {Y.~V.}\ \bibnamefont
  {Kovchegov}},\ }\href {https://doi.org/10.1103/PhysRevD.60.034008} {\bibfield
   {journal} {\bibinfo  {journal} {Phys. Rev. D}\ }\textbf {\bibinfo {volume}
  {60}},\ \bibinfo {pages} {034008} (\bibinfo {year} {1999})},\ \Eprint
  {https://arxiv.org/abs/hep-ph/9901281} {arXiv:hep-ph/9901281} \BibitemShut
  {NoStop}%
\bibitem [{\citenamefont {Collins}\ \emph {et~al.}(1985)\citenamefont
  {Collins}, \citenamefont {Soper},\ and\ \citenamefont
  {Sterman}}]{Collins:1984kg}%
  \BibitemOpen
  \bibfield  {author} {\bibinfo {author} {\bibfnamefont {J.~C.}\ \bibnamefont
  {Collins}}, \bibinfo {author} {\bibfnamefont {D.~E.}\ \bibnamefont {Soper}},\
  and\ \bibinfo {author} {\bibfnamefont {G.~F.}\ \bibnamefont {Sterman}},\
  }\href {https://doi.org/10.1016/0550-3213(85)90479-1} {\bibfield  {journal}
  {\bibinfo  {journal} {Nucl. Phys. B}\ }\textbf {\bibinfo {volume} {250}},\
  \bibinfo {pages} {199} (\bibinfo {year} {1985})}\BibitemShut {NoStop}%
\bibitem [{\citenamefont {Becher}\ and\ \citenamefont
  {Neubert}(2011)}]{Becher:2010tm}%
  \BibitemOpen
  \bibfield  {author} {\bibinfo {author} {\bibfnamefont {T.}~\bibnamefont
  {Becher}}\ and\ \bibinfo {author} {\bibfnamefont {M.}~\bibnamefont
  {Neubert}},\ }\href {https://doi.org/10.1140/epjc/s10052-011-1665-7}
  {\bibfield  {journal} {\bibinfo  {journal} {Eur. Phys. J.}\ }\textbf
  {\bibinfo {volume} {C71}},\ \bibinfo {pages} {1665} (\bibinfo {year}
  {2011})},\ \Eprint {https://arxiv.org/abs/1007.4005} {arXiv:1007.4005
  [hep-ph]} \BibitemShut {NoStop}%
\bibitem [{\citenamefont {Collins}(2011)}]{Collins:2011zzd}%
  \BibitemOpen
  \bibfield  {author} {\bibinfo {author} {\bibfnamefont {J.}~\bibnamefont
  {Collins}},\ }\href {https://doi.org/10.1017/9781009401845} {\emph {\bibinfo
  {title} {{Foundations of Perturbative QCD}}}},\ Vol.~\bibinfo {volume} {32}\
  (\bibinfo  {publisher} {Cambridge University Press},\ \bibinfo {year}
  {2011})\BibitemShut {NoStop}%
\bibitem [{\citenamefont {Catani}\ \emph {et~al.}(2015)\citenamefont {Catani},
  \citenamefont {de~Florian}, \citenamefont {Ferrera},\ and\ \citenamefont
  {Grazzini}}]{Catani:2015vma}%
  \BibitemOpen
  \bibfield  {author} {\bibinfo {author} {\bibfnamefont {S.}~\bibnamefont
  {Catani}}, \bibinfo {author} {\bibfnamefont {D.}~\bibnamefont {de~Florian}},
  \bibinfo {author} {\bibfnamefont {G.}~\bibnamefont {Ferrera}},\ and\ \bibinfo
  {author} {\bibfnamefont {M.}~\bibnamefont {Grazzini}},\ }\href
  {https://doi.org/10.1007/JHEP12(2015)047} {\bibfield  {journal} {\bibinfo
  {journal} {JHEP}\ }\textbf {\bibinfo {volume} {2015}}\bibfield  {number}
  {\bibinfo  {number} { (12)},\ \bibinfo {pages} {047}},\ }\Eprint
  {https://arxiv.org/abs/1507.06937} {arXiv:1507.06937 [hep-ph]} \BibitemShut
  {NoStop}%
\bibitem [{\citenamefont {Sun}\ and\ \citenamefont {Yuan}(2013)}]{Sun:2013hua}%
  \BibitemOpen
  \bibfield  {author} {\bibinfo {author} {\bibfnamefont {P.}~\bibnamefont
  {Sun}}\ and\ \bibinfo {author} {\bibfnamefont {F.}~\bibnamefont {Yuan}},\
  }\href {https://doi.org/10.1103/PhysRevD.88.114012} {\bibfield  {journal}
  {\bibinfo  {journal} {Phys. Rev. D}\ }\textbf {\bibinfo {volume} {88}},\
  \bibinfo {pages} {114012} (\bibinfo {year} {2013})},\ \Eprint
  {https://arxiv.org/abs/1308.5003} {arXiv:1308.5003 [hep-ph]} \BibitemShut
  {NoStop}%
\bibitem [{\citenamefont {Sun}\ \emph {et~al.}(2018)\citenamefont {Sun},
  \citenamefont {Isaacson}, \citenamefont {Yuan},\ and\ \citenamefont
  {Yuan}}]{Sun:2014dqm}%
  \BibitemOpen
  \bibfield  {author} {\bibinfo {author} {\bibfnamefont {P.}~\bibnamefont
  {Sun}}, \bibinfo {author} {\bibfnamefont {J.}~\bibnamefont {Isaacson}},
  \bibinfo {author} {\bibfnamefont {C.~P.}\ \bibnamefont {Yuan}},\ and\
  \bibinfo {author} {\bibfnamefont {F.}~\bibnamefont {Yuan}},\ }\href
  {https://doi.org/10.1142/S0217751X18410063} {\bibfield  {journal} {\bibinfo
  {journal} {Int. J. Mod. Phys. A}\ }\textbf {\bibinfo {volume} {33}},\
  \bibinfo {pages} {1841006} (\bibinfo {year} {2018})},\ \Eprint
  {https://arxiv.org/abs/1406.3073} {arXiv:1406.3073 [hep-ph]} \BibitemShut
  {NoStop}%
\bibitem [{\citenamefont {Collins}\ and\ \citenamefont
  {Rogers}(2015)}]{Collins:2014jpa}%
  \BibitemOpen
  \bibfield  {author} {\bibinfo {author} {\bibfnamefont {J.}~\bibnamefont
  {Collins}}\ and\ \bibinfo {author} {\bibfnamefont {T.}~\bibnamefont
  {Rogers}},\ }\href {https://doi.org/10.1103/PhysRevD.91.074020} {\bibfield
  {journal} {\bibinfo  {journal} {Phys.Rev.}\ }\textbf {\bibinfo {volume}
  {D91}},\ \bibinfo {pages} {074020} (\bibinfo {year} {2015})},\ \Eprint
  {https://arxiv.org/abs/1412.3820} {arXiv:1412.3820 [hep-ph]} \BibitemShut
  {NoStop}%
\bibitem [{\citenamefont {Wei}(2021)}]{Wei:2020glg}%
  \BibitemOpen
  \bibfield  {author} {\bibinfo {author} {\bibfnamefont {S.-y.}\ \bibnamefont
  {Wei}},\ }\href {https://doi.org/10.1016/j.physletb.2021.136356} {\bibfield
  {journal} {\bibinfo  {journal} {Phys. Lett. B}\ }\textbf {\bibinfo {volume}
  {817}},\ \bibinfo {pages} {136356} (\bibinfo {year} {2021})},\ \Eprint
  {https://arxiv.org/abs/2009.06514} {arXiv:2009.06514 [hep-ph]} \BibitemShut
  {NoStop}%
\bibitem [{Sup(2026)}]{SupplementalMaterial}%
  \BibitemOpen
  \href@noop {} {} (\bibinfo {year} {2026}),\ \bibinfo {note} {see Supplemental
  Material for details of the CGC matching and measured recoil,
  finite-rapidity-coverage Sudakov evolution, and its SCET
  derivation}\BibitemShut {NoStop}%
\bibitem [{\citenamefont {Deak}\ \emph {et~al.}(2019)\citenamefont {Deak},
  \citenamefont {van Hameren}, \citenamefont {Jung}, \citenamefont {Kusina},
  \citenamefont {Kutak},\ and\ \citenamefont {Serino}}]{Deak:2018obv}%
  \BibitemOpen
  \bibfield  {author} {\bibinfo {author} {\bibfnamefont {M.}~\bibnamefont
  {Deak}}, \bibinfo {author} {\bibfnamefont {A.}~\bibnamefont {van Hameren}},
  \bibinfo {author} {\bibfnamefont {H.}~\bibnamefont {Jung}}, \bibinfo {author}
  {\bibfnamefont {A.}~\bibnamefont {Kusina}}, \bibinfo {author} {\bibfnamefont
  {K.}~\bibnamefont {Kutak}},\ and\ \bibinfo {author} {\bibfnamefont
  {M.}~\bibnamefont {Serino}},\ }\href
  {https://doi.org/10.1103/PhysRevD.99.094011} {\bibfield  {journal} {\bibinfo
  {journal} {Phys. Rev. D}\ }\textbf {\bibinfo {volume} {99}},\ \bibinfo
  {pages} {094011} (\bibinfo {year} {2019})},\ \Eprint
  {https://arxiv.org/abs/1809.03854} {arXiv:1809.03854 [hep-ph]} \BibitemShut
  {NoStop}%
\bibitem [{\citenamefont {Goncalves}\ \emph {et~al.}(2020)\citenamefont
  {Goncalves}, \citenamefont {Lima}, \citenamefont {Pasechnik},\ and\
  \citenamefont {{\v{S}}umbera}}]{Goncalves:2020tvh}%
  \BibitemOpen
  \bibfield  {author} {\bibinfo {author} {\bibfnamefont {V.~P.}\ \bibnamefont
  {Goncalves}}, \bibinfo {author} {\bibfnamefont {Y.}~\bibnamefont {Lima}},
  \bibinfo {author} {\bibfnamefont {R.}~\bibnamefont {Pasechnik}},\ and\
  \bibinfo {author} {\bibfnamefont {M.}~\bibnamefont {{\v{S}}umbera}},\ }\href
  {https://doi.org/10.1103/PhysRevD.101.094019} {\bibfield  {journal} {\bibinfo
   {journal} {Phys. Rev. D}\ }\textbf {\bibinfo {volume} {101}},\ \bibinfo
  {pages} {094019} (\bibinfo {year} {2020})},\ \Eprint
  {https://arxiv.org/abs/2003.02555} {arXiv:2003.02555 [hep-ph]} \BibitemShut
  {NoStop}%
\bibitem [{\citenamefont {Yang}\ \emph {et~al.}(2022)\citenamefont {Yang} \emph
  {et~al.}}]{Yang:2022qgk}%
  \BibitemOpen
  \bibfield  {author} {\bibinfo {author} {\bibfnamefont {H.}~\bibnamefont
  {Yang}} \emph {et~al.},\ }\href
  {https://doi.org/10.1140/epjc/s10052-022-10715-0} {\bibfield  {journal}
  {\bibinfo  {journal} {Eur. Phys. J. C}\ }\textbf {\bibinfo {volume} {82}},\
  \bibinfo {pages} {755} (\bibinfo {year} {2022})},\ \Eprint
  {https://arxiv.org/abs/2204.01528} {arXiv:2204.01528 [hep-ph]} \BibitemShut
  {NoStop}%
\bibitem [{\citenamefont {Ganguli}\ \emph {et~al.}(2023)\citenamefont
  {Ganguli}, \citenamefont {van Hameren}, \citenamefont {Kotko},\ and\
  \citenamefont {Kutak}}]{Ganguli:2023joy}%
  \BibitemOpen
  \bibfield  {author} {\bibinfo {author} {\bibfnamefont {I.}~\bibnamefont
  {Ganguli}}, \bibinfo {author} {\bibfnamefont {A.}~\bibnamefont {van
  Hameren}}, \bibinfo {author} {\bibfnamefont {P.}~\bibnamefont {Kotko}},\ and\
  \bibinfo {author} {\bibfnamefont {K.}~\bibnamefont {Kutak}},\ }\href
  {https://doi.org/10.1140/epjc/s10052-023-12043-3} {\bibfield  {journal}
  {\bibinfo  {journal} {Eur. Phys. J. C}\ }\textbf {\bibinfo {volume} {83}},\
  \bibinfo {pages} {868} (\bibinfo {year} {2023})},\ \Eprint
  {https://arxiv.org/abs/2306.04706} {arXiv:2306.04706 [hep-ph]} \BibitemShut
  {NoStop}%
\bibitem [{\citenamefont {Taels}(2024)}]{Taels:2023czt}%
  \BibitemOpen
  \bibfield  {author} {\bibinfo {author} {\bibfnamefont {P.}~\bibnamefont
  {Taels}},\ }\href {https://doi.org/10.1007/JHEP01(2024)005} {\bibfield
  {journal} {\bibinfo  {journal} {JHEP}\ }\textbf {\bibinfo {volume}
  {2024}}\bibfield  {number} {\bibinfo  {number} { (01)},\ \bibinfo {pages}
  {005}},\ }\Eprint {https://arxiv.org/abs/2308.02449} {arXiv:2308.02449
  [hep-ph]} \BibitemShut {NoStop}%
\bibitem [{\citenamefont {Bandeira}\ \emph {et~al.}(2025)\citenamefont
  {Bandeira}, \citenamefont {Goncalves},\ and\ \citenamefont
  {Sch{\"a}fer}}]{Bandeira:2024jjl}%
  \BibitemOpen
  \bibfield  {author} {\bibinfo {author} {\bibfnamefont {Y.~B.}\ \bibnamefont
  {Bandeira}}, \bibinfo {author} {\bibfnamefont {V.~P.}\ \bibnamefont
  {Goncalves}},\ and\ \bibinfo {author} {\bibfnamefont {W.}~\bibnamefont
  {Sch{\"a}fer}},\ }\href {https://doi.org/10.1103/PhysRevD.111.074041}
  {\bibfield  {journal} {\bibinfo  {journal} {Phys. Rev. D}\ }\textbf {\bibinfo
  {volume} {111}},\ \bibinfo {pages} {074041} (\bibinfo {year} {2025})},\
  \Eprint {https://arxiv.org/abs/2411.12675} {arXiv:2411.12675 [hep-ph]}
  \BibitemShut {NoStop}%
\bibitem [{\citenamefont {Albacete}\ \emph {et~al.}(2011)\citenamefont
  {Albacete}, \citenamefont {Armesto}, \citenamefont {Milhano}, \citenamefont
  {Quiroga-Arias},\ and\ \citenamefont {Salgado}}]{Albacete:2010sy}%
  \BibitemOpen
  \bibfield  {author} {\bibinfo {author} {\bibfnamefont {J.~L.}\ \bibnamefont
  {Albacete}}, \bibinfo {author} {\bibfnamefont {N.}~\bibnamefont {Armesto}},
  \bibinfo {author} {\bibfnamefont {J.~G.}\ \bibnamefont {Milhano}}, \bibinfo
  {author} {\bibfnamefont {P.}~\bibnamefont {Quiroga-Arias}},\ and\ \bibinfo
  {author} {\bibfnamefont {C.~A.}\ \bibnamefont {Salgado}},\ }\href
  {https://doi.org/10.1140/epjc/s10052-011-1705-3} {\bibfield  {journal}
  {\bibinfo  {journal} {Eur. Phys. J. C}\ }\textbf {\bibinfo {volume} {71}},\
  \bibinfo {pages} {1705} (\bibinfo {year} {2011})},\ \Eprint
  {https://arxiv.org/abs/1012.4408} {arXiv:1012.4408 [hep-ph]} \BibitemShut
  {NoStop}%
\bibitem [{\citenamefont {Chiu}\ \emph {et~al.}(2012)\citenamefont {Chiu},
  \citenamefont {Jain}, \citenamefont {Neill},\ and\ \citenamefont
  {Rothstein}}]{Chiu:2012ir}%
  \BibitemOpen
  \bibfield  {author} {\bibinfo {author} {\bibfnamefont {J.-Y.}\ \bibnamefont
  {Chiu}}, \bibinfo {author} {\bibfnamefont {A.}~\bibnamefont {Jain}}, \bibinfo
  {author} {\bibfnamefont {D.}~\bibnamefont {Neill}},\ and\ \bibinfo {author}
  {\bibfnamefont {I.~Z.}\ \bibnamefont {Rothstein}},\ }\href
  {https://doi.org/10.1007/JHEP05(2012)084} {\bibfield  {journal} {\bibinfo
  {journal} {JHEP}\ }\textbf {\bibinfo {volume} {2012}}\bibfield  {number}
  {\bibinfo  {number} { (05)},\ \bibinfo {pages} {084}},\ }\Eprint
  {https://arxiv.org/abs/1202.0814} {arXiv:1202.0814 [hep-ph]} \BibitemShut
  {NoStop}%
\bibitem [{\citenamefont {Dasgupta}\ and\ \citenamefont
  {Salam}(2001)}]{Dasgupta:2001sh}%
  \BibitemOpen
  \bibfield  {author} {\bibinfo {author} {\bibfnamefont {M.}~\bibnamefont
  {Dasgupta}}\ and\ \bibinfo {author} {\bibfnamefont {G.~P.}\ \bibnamefont
  {Salam}},\ }\href {https://doi.org/10.1016/S0370-2693(01)00725-0} {\bibfield
  {journal} {\bibinfo  {journal} {Phys. Lett.}\ }\textbf {\bibinfo {volume}
  {B512}},\ \bibinfo {pages} {323} (\bibinfo {year} {2001})},\ \Eprint
  {https://arxiv.org/abs/hep-ph/0104277} {arXiv:hep-ph/0104277 [hep-ph]}
  \BibitemShut {NoStop}%
\bibitem [{\citenamefont {Banfi}\ \emph {et~al.}(2002)\citenamefont {Banfi},
  \citenamefont {Marchesini},\ and\ \citenamefont {Smye}}]{Banfi:2002hw}%
  \BibitemOpen
  \bibfield  {author} {\bibinfo {author} {\bibfnamefont {A.}~\bibnamefont
  {Banfi}}, \bibinfo {author} {\bibfnamefont {G.}~\bibnamefont {Marchesini}},\
  and\ \bibinfo {author} {\bibfnamefont {G.}~\bibnamefont {Smye}},\ }\href
  {https://doi.org/10.1088/1126-6708/2002/08/006} {\bibfield  {journal}
  {\bibinfo  {journal} {JHEP}\ }\textbf {\bibinfo {volume} {2002}}\bibfield
  {number} {\bibinfo  {number} { (08)},\ \bibinfo {pages} {006}},\ }\Eprint
  {https://arxiv.org/abs/hep-ph/0206076} {arXiv:hep-ph/0206076 [hep-ph]}
  \BibitemShut {NoStop}%
\bibitem [{\citenamefont {Becher}\ \emph
  {et~al.}(2016{\natexlab{a}})\citenamefont {Becher}, \citenamefont {Neubert},
  \citenamefont {Rothen},\ and\ \citenamefont {Shao}}]{Becher:2015hka}%
  \BibitemOpen
  \bibfield  {author} {\bibinfo {author} {\bibfnamefont {T.}~\bibnamefont
  {Becher}}, \bibinfo {author} {\bibfnamefont {M.}~\bibnamefont {Neubert}},
  \bibinfo {author} {\bibfnamefont {L.}~\bibnamefont {Rothen}},\ and\ \bibinfo
  {author} {\bibfnamefont {D.~Y.}\ \bibnamefont {Shao}},\ }\href
  {https://doi.org/10.1103/PhysRevLett.116.192001} {\bibfield  {journal}
  {\bibinfo  {journal} {Phys. Rev. Lett.}\ }\textbf {\bibinfo {volume} {116}},\
  \bibinfo {pages} {192001} (\bibinfo {year} {2016}{\natexlab{a}})},\ \Eprint
  {https://arxiv.org/abs/1508.06645} {arXiv:1508.06645 [hep-ph]} \BibitemShut
  {NoStop}%
\bibitem [{\citenamefont {Larkoski}\ \emph {et~al.}(2015)\citenamefont
  {Larkoski}, \citenamefont {Moult},\ and\ \citenamefont
  {Neill}}]{Larkoski:2015zka}%
  \BibitemOpen
  \bibfield  {author} {\bibinfo {author} {\bibfnamefont {A.~J.}\ \bibnamefont
  {Larkoski}}, \bibinfo {author} {\bibfnamefont {I.}~\bibnamefont {Moult}},\
  and\ \bibinfo {author} {\bibfnamefont {D.}~\bibnamefont {Neill}},\ }\href
  {https://doi.org/10.1007/JHEP09(2015)143} {\bibfield  {journal} {\bibinfo
  {journal} {JHEP}\ }\textbf {\bibinfo {volume} {09}},\ \bibinfo {pages}
  {143}},\ \Eprint {https://arxiv.org/abs/1501.04596} {arXiv:1501.04596
  [hep-ph]} \BibitemShut {NoStop}%
\bibitem [{\citenamefont {Caron-Huot}(2018)}]{Caron-Huot:2015bja}%
  \BibitemOpen
  \bibfield  {author} {\bibinfo {author} {\bibfnamefont {S.}~\bibnamefont
  {Caron-Huot}},\ }\href {https://doi.org/10.1007/JHEP03(2018)036} {\bibfield
  {journal} {\bibinfo  {journal} {JHEP}\ }\textbf {\bibinfo {volume} {03}},\
  \bibinfo {pages} {036}},\ \Eprint {https://arxiv.org/abs/1501.03754}
  {arXiv:1501.03754 [hep-ph]} \BibitemShut {NoStop}%
\bibitem [{\citenamefont {Becher}\ \emph
  {et~al.}(2016{\natexlab{b}})\citenamefont {Becher}, \citenamefont {Neubert},
  \citenamefont {Rothen},\ and\ \citenamefont {Shao}}]{Becher:2016mmh}%
  \BibitemOpen
  \bibfield  {author} {\bibinfo {author} {\bibfnamefont {T.}~\bibnamefont
  {Becher}}, \bibinfo {author} {\bibfnamefont {M.}~\bibnamefont {Neubert}},
  \bibinfo {author} {\bibfnamefont {L.}~\bibnamefont {Rothen}},\ and\ \bibinfo
  {author} {\bibfnamefont {D.~Y.}\ \bibnamefont {Shao}},\ }\href
  {https://doi.org/10.1007/JHEP11(2016)019, 10.1007/JHEP05(2017)154} {\bibfield
   {journal} {\bibinfo  {journal} {JHEP}\ }\textbf {\bibinfo {volume} {11}},\
  \bibinfo {pages} {019}},\ \bibinfo {note} {[Erratum: JHEP05,154(2017)]},\
  \Eprint {https://arxiv.org/abs/1605.02737} {arXiv:1605.02737 [hep-ph]}
  \BibitemShut {NoStop}%
\bibitem [{\citenamefont {Becher}\ \emph
  {et~al.}(2016{\natexlab{c}})\citenamefont {Becher}, \citenamefont {Pecjak},\
  and\ \citenamefont {Shao}}]{Becher:2016omr}%
  \BibitemOpen
  \bibfield  {author} {\bibinfo {author} {\bibfnamefont {T.}~\bibnamefont
  {Becher}}, \bibinfo {author} {\bibfnamefont {B.~D.}\ \bibnamefont {Pecjak}},\
  and\ \bibinfo {author} {\bibfnamefont {D.~Y.}\ \bibnamefont {Shao}},\ }\href
  {https://doi.org/10.1007/JHEP12(2016)018} {\bibfield  {journal} {\bibinfo
  {journal} {JHEP}\ }\textbf {\bibinfo {volume} {2016}}\bibfield  {number}
  {\bibinfo  {number} { (12)},\ \bibinfo {pages} {018}},\ }\Eprint
  {https://arxiv.org/abs/1610.01608} {arXiv:1610.01608 [hep-ph]} \BibitemShut
  {NoStop}%
\bibitem [{\citenamefont {Forshaw}\ \emph {et~al.}(2006)\citenamefont
  {Forshaw}, \citenamefont {Kyrieleis},\ and\ \citenamefont
  {Seymour}}]{Forshaw:2006fk}%
  \BibitemOpen
  \bibfield  {author} {\bibinfo {author} {\bibfnamefont {J.~R.}\ \bibnamefont
  {Forshaw}}, \bibinfo {author} {\bibfnamefont {A.}~\bibnamefont {Kyrieleis}},\
  and\ \bibinfo {author} {\bibfnamefont {M.~H.}\ \bibnamefont {Seymour}},\
  }\href {https://doi.org/10.1088/1126-6708/2006/08/059} {\bibfield  {journal}
  {\bibinfo  {journal} {JHEP}\ }\textbf {\bibinfo {volume} {2006}}\bibfield
  {number} {\bibinfo  {number} { (08)},\ \bibinfo {pages} {059}},\ }\Eprint
  {https://arxiv.org/abs/hep-ph/0604094} {arXiv:hep-ph/0604094} \BibitemShut
  {NoStop}%
\bibitem [{\citenamefont {Gaunt}(2014)}]{Gaunt:2014ska}%
  \BibitemOpen
  \bibfield  {author} {\bibinfo {author} {\bibfnamefont {J.~R.}\ \bibnamefont
  {Gaunt}},\ }\href {https://doi.org/10.1007/JHEP07(2014)110} {\bibfield
  {journal} {\bibinfo  {journal} {JHEP}\ }\textbf {\bibinfo {volume}
  {2014}}\bibfield  {number} {\bibinfo  {number} { (07)},\ \bibinfo {pages}
  {110}},\ }\Eprint {https://arxiv.org/abs/1405.2080} {arXiv:1405.2080
  [hep-ph]} \BibitemShut {NoStop}%
\bibitem [{\citenamefont {Rothstein}\ and\ \citenamefont
  {Stewart}(2016)}]{Rothstein:2016bsq}%
  \BibitemOpen
  \bibfield  {author} {\bibinfo {author} {\bibfnamefont {I.~Z.}\ \bibnamefont
  {Rothstein}}\ and\ \bibinfo {author} {\bibfnamefont {I.~W.}\ \bibnamefont
  {Stewart}},\ }\href {https://doi.org/10.1007/JHEP08(2016)025} {\bibfield
  {journal} {\bibinfo  {journal} {JHEP}\ }\textbf {\bibinfo {volume}
  {2016}}\bibfield  {number} {\bibinfo  {number} { (08)},\ \bibinfo {pages}
  {025}},\ }\Eprint {https://arxiv.org/abs/1601.04695} {arXiv:1601.04695
  [hep-ph]} \BibitemShut {NoStop}%
\bibitem [{\citenamefont {Becher}\ \emph {et~al.}(2021)\citenamefont {Becher},
  \citenamefont {Neubert},\ and\ \citenamefont {Shao}}]{Becher:2021zkk}%
  \BibitemOpen
  \bibfield  {author} {\bibinfo {author} {\bibfnamefont {T.}~\bibnamefont
  {Becher}}, \bibinfo {author} {\bibfnamefont {M.}~\bibnamefont {Neubert}},\
  and\ \bibinfo {author} {\bibfnamefont {D.~Y.}\ \bibnamefont {Shao}},\ }\href
  {https://doi.org/10.1103/PhysRevLett.127.212002} {\bibfield  {journal}
  {\bibinfo  {journal} {Phys. Rev. Lett.}\ }\textbf {\bibinfo {volume} {127}},\
  \bibinfo {pages} {212002} (\bibinfo {year} {2021})},\ \Eprint
  {https://arxiv.org/abs/2107.01212} {arXiv:2107.01212 [hep-ph]} \BibitemShut
  {NoStop}%
\bibitem [{\citenamefont {Banfi}\ \emph {et~al.}(2026)\citenamefont {Banfi},
  \citenamefont {Forshaw},\ and\ \citenamefont {Holguin}}]{Banfi:2025mra}%
  \BibitemOpen
  \bibfield  {author} {\bibinfo {author} {\bibfnamefont {A.}~\bibnamefont
  {Banfi}}, \bibinfo {author} {\bibfnamefont {J.~R.}\ \bibnamefont {Forshaw}},\
  and\ \bibinfo {author} {\bibfnamefont {J.}~\bibnamefont {Holguin}},\ }\href
  {https://doi.org/10.1103/flv7-ksvf} {\bibfield  {journal} {\bibinfo
  {journal} {Phys. Rev. Lett.}\ }\textbf {\bibinfo {volume} {136}},\ \bibinfo
  {pages} {221901} (\bibinfo {year} {2026})},\ \Eprint
  {https://arxiv.org/abs/2511.11799} {arXiv:2511.11799 [hep-ph]} \BibitemShut
  {NoStop}%
\bibitem [{\citenamefont {Becher}\ \emph {et~al.}(2026)\citenamefont {Becher},
  \citenamefont {Hager}, \citenamefont {Neubert},\ and\ \citenamefont
  {Schwienbacher}}]{Becher:2026kbr}%
  \BibitemOpen
  \bibfield  {author} {\bibinfo {author} {\bibfnamefont {T.}~\bibnamefont
  {Becher}}, \bibinfo {author} {\bibfnamefont {P.}~\bibnamefont {Hager}},
  \bibinfo {author} {\bibfnamefont {M.}~\bibnamefont {Neubert}},\ and\ \bibinfo
  {author} {\bibfnamefont {D.}~\bibnamefont {Schwienbacher}},\ }\href@noop {}
  {\bibinfo {title} {{Factorization Beyond Coherence}}} (\bibinfo {year}
  {2026}),\ \Eprint {https://arxiv.org/abs/2603.12383} {arXiv:2603.12383
  [hep-ph]} \BibitemShut {NoStop}%
\bibitem [{\citenamefont {Barcaro}\ \emph {et~al.}(2026)\citenamefont
  {Barcaro}, \citenamefont {Gao}, \citenamefont {Gaunt},\ and\ \citenamefont
  {Pathak}}]{Barcaro:2026dsd}%
  \BibitemOpen
  \bibfield  {author} {\bibinfo {author} {\bibfnamefont {D.}~\bibnamefont
  {Barcaro}}, \bibinfo {author} {\bibfnamefont {A.}~\bibnamefont {Gao}},
  \bibinfo {author} {\bibfnamefont {J.~R.}\ \bibnamefont {Gaunt}},\ and\
  \bibinfo {author} {\bibfnamefont {A.}~\bibnamefont {Pathak}},\ }\href@noop {}
  {\bibinfo {title} {{Collinear Factorization Violation and Reggeization}}}
  (\bibinfo {year} {2026}),\ \Eprint {https://arxiv.org/abs/2607.14259}
  {arXiv:2607.14259 [hep-ph]} \BibitemShut {NoStop}%
\end{thebibliography}%
	
	\clearpage
	\appendix
	\onecolumngrid
	
	\makeatletter
	\@removefromreset{equation}{section}
	\makeatother
	
	\setcounter{equation}{0}
	\setcounter{figure}{0}
	\setcounter{table}{0}
	\renewcommand{\theequation}{S-\arabic{equation}}
	\renewcommand{\thefigure}{S-\arabic{figure}}
	\renewcommand{\thetable}{S-\arabic{table}}
	
	\renewcommand{\theHequation}{supp.equation.\arabic{equation}}
	\renewcommand{\theHfigure}{supp.figure.\arabic{figure}}
	\renewcommand{\theHtable}{supp.table.\arabic{table}}
	
	\allowdisplaybreaks
	
	\section*{Supplemental Material} 
	
	\subsection{CGC matching and the measured recoil}
	The acceptance-dependent target beam function in Eq.~\eqref{eq:target_beam_matching} requires the $g\to q\bar q$ small-$x$ splitting to be kept differential in the companion momentum. 
	This pair-differential matching exposes the transverse momentum delivered by the nuclear field before Drell--Yan evolution broadens the observed recoil.  
	For on-shell $Z^0$ production, we use
	\begin{equation}
		Q=M_Z,\qquad m_T=\sqrt{Q^2+q_T^2},\qquad
		x_p=\frac{m_T e^{y_Z}}{\sqrt{s_{NN}}},\qquad
		x_A=\frac{m_T e^{-y_Z}}{\sqrt{s_{NN}}},\qquad
		\sigma_{0,q}^{Z}=\frac{\sqrt{2}\pi G_F M_Z^2}{N_c s_{NN}}\left(V_q^2+A_q^2\right).
		\label{app:born_kinematics}
	\end{equation}
	Here $V_q=T_3^q-2Q_q\sin^2\theta_W$ and $A_q=T_3^q$; the tree-level hard coefficient is set to unity.  
	Forward rapidity gives $x_p/x_A=e^{2y_Z}$: the proton is probed in its dilute quark sector while the nucleus is probed in its small-$x$ gluon field.  
	The scale $Q$ fixes the short annihilation time; the nuclear transverse structure lies at the semihard scale $Q_s$.
	
	The matching is most transparent in the leading-power region $Q^2\gg k_{\bar qT}^2, Q_s^2$.  
	The target supplies transverse momentum $\boldsymbol\kappa_T$ to the last $g\to q\bar q$ pair; if the companion carries $\boldsymbol p_T$, the antiquark entering the hard collision carries $\boldsymbol k_{\bar q}=\boldsymbol\kappa_T-\boldsymbol p_T$.  
	At its natural TMD scale, the dipole correlator fixes the pair density directly~\cite{Marquet:2009ca, Xiao:2010sa}:
	\begin{align}
		x_A\mathcal P_{\bar q q/A} 	(x_A;\xi,\boldsymbol\kappa_T,\boldsymbol p_T) ={}&\frac{N_c\mathcal A_T}{8\pi^4} 
		\int_0^\infty\frac{r_T\,dr_T}{2\pi} J_0(\kappa_T r_T)S_A\!\left(r_T,\frac{x_A}{\xi}\right)
		\left| 	\frac{(\boldsymbol\kappa_T-\boldsymbol p_T)p_T} {(1-\xi)|\boldsymbol\kappa_T-\boldsymbol p_T|^2 +\xi p_T^2}
		+\frac{\boldsymbol p_T}{p_T} \right|^2.
		\label{app:cgc_matching}
	\end{align}
	Here $x_g$ is the longitudinal momentum fraction of the parent gluon, and $\xi=x_A/x_g$ is the fraction carried by the active antiquark.  
	We use $p_T=|\boldsymbol p_T|$ and $\kappa_T=|\boldsymbol\kappa_T|$, while $r_T$ denotes the dipole size.  
	The quantity $N_A(r_T,x_g)$ is the dipole scattering amplitude for target $A$, $S_A=1-N_A$, and $\mathcal A_T$ denotes the effective transverse area.
	The two terms inside the modulus arise from the splitting wave function and its eikonal gauge-link contribution.  
	The dipole distribution fixes the total momentum transfer $\boldsymbol\kappa_T$, while the splitting kernel determines how this momentum is shared between the active antiquark and its companion.
	A dipole of size $r_T$ samples momenta of order $1/r_T$, so the growth of $N_A$ near $r_T\sim1/Q_s$ broadens the $\kappa_T$ distribution over $\kappa_T\sim Q_s$. 
	Its small-$x$ evolution is the running coupling BK (rcBK) evolution of $N_A$~\cite{Albacete:2010sy}.  
	The proton, pA2, and pA3 inputs differ only in their initial saturation scale and use the common area $\mathcal A_T=15~\mathrm{mb}$~\cite{Marquet:2019ltn}.
	In the phenomenological calculations, we solve the rcBK equation using the prescription of Ref.~\cite{Albacete:2010sy}, with the MV initial condition~\cite{McLerran:1993ni,McLerran:1993ka}
	\begin{equation}
		N_p(r)=1-\exp\left[-\frac{1}{4}Q_{s0,\,p}^2\,r^2\ln\left(e+\frac{1}{r\Lambda}\right)\right].
	\end{equation}
	where $Q_{s0,p}^2=0.2~\mathrm{GeV}^2$ and $\Lambda=0.24~\mathrm{GeV}$.
	
	Integrating the pair density over the companion momentum yields the inclusive target beam function,
	\begin{equation}
		\mathcal B^{\rm inc}_{\bar q/A}(x_A,\boldsymbol k_{\bar qT})
		=\int_{x_A}^{1}d\xi\int d^2p_T\, \mathcal P_{\bar q q/A} (x_A;\xi,\boldsymbol k_{\bar qT}+\boldsymbol p_T, \boldsymbol p_T).
		\label{app:sea_tmd}
	\end{equation}
	Thus $[\mathcal P_{\bar q q/A}]={\rm GeV}^{-4}$, $[\mathcal B^{\rm inc}_{\bar q/A}(\boldsymbol k_{\bar qT})] ={\rm GeV}^{-2}$, and the Fourier-space beam function is dimensionless. 
	This integral gives the usual small-$x$ sea distribution: it is saturated for $k_{\bar qT}\lesssim Q_s$ and has the leading-twist $Q_s^2/k_{\bar qT}^2$ tail.  
	It also removes the direction of $\boldsymbol\kappa_T$.  
	By imposing the coverage measurement before the companion-momentum integration, the acceptance-dependent target beam function retains information on whether the companion recoil lies inside $\mathcal C$.
	For the massless companion quark in consideration, we have
	\begin{equation}
		\eta_{\rm comp}^{\rm lab} =y_Z^{\rm lab}+\ln\frac{\xi p_T}{(1-\xi)Q},
		\qquad 	\chi_{\mathcal C}(\eta_{\rm comp}^{\rm lab})= \Theta(\eta_{\rm comp}^{\rm lab}-\eta_{\min}^{\rm lab})\, \Theta(\eta_{\max}^{\rm lab}-\eta_{\rm comp}^{\rm lab}).
		\label{app:companion_coverage}
	\end{equation}
	Inserting this coverage function in Eq.~\eqref{eq:target_beam_matching} applies the fiducial measurement to the partonic matching state, while the observable remains the inclusive hadronic sum in Eq.~\eqref{eq:vector_sum}, not a quark tag. 
	At $\boldsymbol b_k=\boldsymbol b$ and $\boldsymbol b_l=-\boldsymbol b$, the two measurement branches acquire the same Fourier phase, and the target projection reduces to
	\begin{align}
		\left.\widetilde{\mathcal B}^{\mathcal C}_{\bar q/A}(x_A,\boldsymbol b,-\boldsymbol b)\right|_{\rm CGC}
		={}&\int_{x_A}^{1}d\xi\int d^2\kappa_T\,d^2p_T\, e^{-i\boldsymbol b\cdot (\boldsymbol\kappa_T-\boldsymbol p_T)} \mathcal P_{\bar q q/A} (x_A;\xi,\boldsymbol\kappa_T,\boldsymbol p_T) 	\nonumber\\
		={}&\int d^2k_{\bar qT}\, e^{-i\boldsymbol b\cdot\boldsymbol k_{\bar qT}} \mathcal B^{\rm inc}_{\bar q/A}(x_A,\boldsymbol k_{\bar qT}).
		\label{app:beam_inclusive_limit}
	\end{align}
	Thus $\mathcal B^{\rm inc}_{\bar q/A}$ and $\widetilde{\mathcal B}^{\mathcal C}_{\bar q/A}$ are two projections of the same pair density: the former integrates inclusively over the companion, whereas the latter retains whether its recoil is measured inside $\mathcal C$.
	Covered incoming radiation remains in $R_{\rm cov}^{\rm pert}$.  
	Rotational invariance converts the joint vector density to the recoil-angle density with the Jacobian $4\pi q_T l_T$ after setting $\boldsymbol q_T\!\cdot\!\boldsymbol l_T =q_T l_T\cos\Delta\phi_{Zl}$.
	
	A finite vector-angularity cut acts only after the measured vector has been formed. 
	It restricts the magnitude of the summed measured recoil to the disk $l_T<\tau_0$, rather than vetoing each emission separately.  
	For the $q_T$ spectrum this integration holds $\boldsymbol q_T$ fixed, so $\boldsymbol k_T=\boldsymbol q_T+\boldsymbol l_T$ varies with $\boldsymbol l_T$ and the Fourier transform couples the two coordinates:
	\begin{equation}
		\int_{l_T<\tau_0}d^2l_T\, e^{i(\boldsymbol b_l+\boldsymbol b_k)\cdot\boldsymbol l_T} =2\pi\tau_0\, \frac{J_1\!\left(\tau_0|\boldsymbol b_l+\boldsymbol b_k|\right)}{|\boldsymbol b_l+\boldsymbol b_k|}.
		\label{app:veto_kernel}
	\end{equation}
	For $\tau_0\to\infty$ this becomes $(2\pi)^2\delta^{(2)}(\boldsymbol b_l+\boldsymbol b_k)$.  
	The two Fourier coordinates then become opposite, and the target beam function reduces to its inclusive boundary.  
	The full inclusive limit is verified below.
	
	\subsection{Sudakov evolution with finite rapidity coverage}
	
	At small $q_T$, the Drell--Yan logarithm counts soft radiation over both transverse scale and rapidity.  
	The hard annihilation fixes the ends of this phase space, while the detector divides it according to where the recoil is recorded.
	The rcBK-evolved dipole amplitude encodes the small-$x$ evolution of the target gluon field prior to the last $g\to q\bar q$ splitting.  
	This splitting matches the dipole distribution onto the target sea-antiquark TMD.  
	From their natural transverse scales to $Q$, the target antiquark and projectile quark TMDs undergo the usual Drell--Yan Sudakov evolution.  
	At one loop, the combined evolution of the two incoming quark legs yields the standard inclusive Sudakov factor,
	\begin{equation}
		R_{\rm DY}^{\rm pert}(b,Q) =\frac{C_F}{\pi}\int_{\mu_b^2}^{Q^2}\frac{d\mu^2}{\mu^2}\, \alpha_s(\mu)\left[\ln\frac{Q^2}{\mu^2}-\frac32\right].
		\label{app:inclusive_sudakov}
	\end{equation}
	Here $C_F=(N_c^2-1)/(2N_c)$, and $\mu_b$ is the natural TMD scale. 
	The logarithm is the cusp contribution; the finite $-3/2$ term belongs to the collinear beam evolution.  
	Only the former carries a rapidity length.  
	Since $dy_s=dl^+/l^+$, a soft gluon of transverse scale $\mu$ contributes
	\begin{equation}
		dP_{\rm soft}=\frac{C_F}{\pi}\alpha_s(\mu) \frac{d\mu^2}{\mu^2}\,dy_s,
		\qquad |y_s-y_Z|<L_\mu,
		\qquad L_\mu\equiv\ln\frac{Q}{\mu}.
		\label{app:soft_phase_space}
	\end{equation}
	The full interval has length $2L_\mu=\ln(Q^2/\mu^2)$.  
	The overlap in Eq.~\eqref{eq:covered_rapidity} is narrow near the hard scale, opens as $\mu$ decreases, and eventually contains the full acceptance. 
	Real--virtual cancellation gives Eqs.~\eqref{eq:covered_radiator} and \eqref{eq:residual_radiator}. 
	The first Sudakov factor $R_{\rm cov}^{\rm pert}$ accompanies $\boldsymbol l_T$; the second $R_{\rm rest}^{\rm pert}$ accompanies $\boldsymbol q_T+\boldsymbol l_T$.  
	The beam term has no rapidity length and therefore remains in the latter.  
	Their sum is the inclusive Drell--Yan Sudakov factor.
	
	Once the soft interval contains the full detector coverage, $\Delta\eta_{\rm cov}=\Delta\eta_{\mathcal C}^{\rm lab}=\eta_{\max}^{\rm lab}-\eta_{\min}^{\rm lab}$, and the detector removes a fixed length $\Delta\eta_{\mathcal C}^{\rm lab}$ from $\int dl^+/l^+$.
	The two uncovered ends combine into one logarithm, $2L_\mu-\Delta\eta_{\mathcal C}^{\rm lab} =\ln(Q_{\rm eff}^2/\mu^2)$, with $Q_{\rm eff}$ defined in Eq.~\eqref{eq:effective_sudakov_scale}.
	For a symmetric interval $|y_s-y_Z|<y_0$, this is the rapidity split obtained directly from the real--virtual integral in the heuristic derivation, with $Q_{\rm eff}=Qe^{-y_0}$.  
	Keeping the two measured Fourier coordinates independent gives
	\begin{align}
		R_{\rm rest}^{\rm pert}(b_k) ={}&\frac{C_F}{\pi}\left[ \int_{\mu_{b_k}^2}^{Q_{\rm eff}^2}\frac{d\mu^2}{\mu^2}\,
		\alpha_s(\mu)\ln\frac{Q_{\rm eff}^2}{\mu^2} -\frac32\int_{\mu_{b_k}^2}^{Q^2}\frac{d\mu^2}{\mu^2}\, \alpha_s(\mu)\right], \nonumber\\
		R_{\rm cov}^{\rm pert}(b_l) ={}&\frac{C_F}{\pi}\left[ \int_{\mu_{b_l}^2}^{Q^2}\frac{d\mu^2}{\mu^2}\,
		\alpha_s(\mu)\ln\frac{Q^2}{\mu^2} -\int_{\mu_{b_l}^2}^{Q_{\rm eff}^2}\frac{d\mu^2}{\mu^2}\, \alpha_s(\mu)\ln\frac{Q_{\rm eff}^2}{\mu^2} \right].
		\label{app:heuristic_split}
	\end{align}
	The $Q_{\rm eff}$ integrals stop when $\mu_b$ reaches $Q_{\rm eff}$. 
	Eqs.~\eqref{eq:covered_rapidity}--\eqref{eq:residual_radiator} retain the exact edge turn-on for the asymmetric laboratory interval used in the numerical calculation; Eq.~\eqref{app:heuristic_split} displays the momentum flow.  
	The first line is the radiation that survives the fiducial vector cancellation, with cusp evolution ending at $Q_{\rm eff}$.  
	The second line is the complementary cusp phase space carried by the measured recoil.
	
	For the numerical Fourier transforms we use the standard $b_*$ profile and nonperturbative form~\cite{Sun:2013hua, Sun:2014dqm, Collins:2014jpa, Wei:2020glg}:
	\begin{align}
		b_*&=\frac{b}{\sqrt{1+b^2/b_{\max}^2}},
		\qquad \mu_b=\min\!\left(\frac{2e^{-\gamma_E}}{b_*},Q\right), \nonumber\\
		R_{\rm NP}(b,Q) &=0.106~{\rm GeV}^2b^2 +0.42\ln\frac{Q}{Q_0}\ln\frac{b}{b_*},
		\qquad Q_0=\sqrt{2.4}~{\rm GeV},
		\label{app:profile_and_np}
	\end{align}
	with $b_{\max}=1.5~{\rm GeV}^{-1}$.
	The $b_*$ map holds the perturbative scale above the infrared region, while $R_{\rm NP}$ supplies the continuing large-$b$ broadening. 
	The nonperturbative transverse-momentum dynamics is modeled differently on the two sides: target-side broadening is encoded by saturation at the semihard scale $Q_s$, whereas the projectile-side contribution is described by $R_{\rm NP}$.
	The target is, instead, governed by the saturation physics with a semihard scale $Q_s$. 
	Consequently, we include only half of the Drell--Yan nonperturbative Sudakov factor in \cref{app:profile_and_np} to avoid double counting.
	The nonperturbative Sudakov factor is assigned to $R_{\rm rest}=R_{\rm rest}^{\rm pert}+R_{\rm NP}$: it describes the intrinsic and long-distance transverse momentum that remains in the residual vector. 
	The perturbative Sudakov factors use one-loop running with $N_f=4$ and $\Lambda_{\rm QCD}=0.155~{\rm GeV}$.
	
	Integrating over the measured recoil sets the two impact-parameter vectors to opposite values with equal magnitudes. 
	Eqs.~\eqref{eq:covered_radiator} and \eqref{eq:residual_radiator} recover the inclusive Drell--Yan Sudakov factor stated below Eq.~\eqref{eq:joint_cross_section}.
	Together with the beam-function identity in Eq.~\eqref{app:beam_inclusive_limit}, this gives $\int d^2l_T\,d\sigma_A/(dy_Z\,d^2q_T\,d^2l_T) =d\sigma_A^{\rm inc}/(dy_Z\,d^2q_T)$.
	
	\subsection{SCET derivation of the finite-coverage Sudakov evolution}
	
	%
	For $q_T\ll Q$ and a fixed central coverage, the beam collinear modes have rapidities of order $\ln(Q/q_T)$ and lie outside the measured interval at leading power.  
	They therefore enter the residual recoil; the covered radiative recoil is soft.  
	Target-fragmentation energy flow is kept through the CGC beam matching above.
	
	Before this matching, let $\boldsymbol k_n$ and $\boldsymbol k_{\bar n}$ be the two collinear transverse momenta entering the hard collision, and let $\boldsymbol r_o$ and $\boldsymbol r_i$ be the physical final-state soft-momentum sums outside and inside the coverage.  
	Momentum conservation gives $\boldsymbol q_T=\boldsymbol k_n+\boldsymbol k_{\bar n} -\boldsymbol r_o-\boldsymbol r_i$.  
	Since $\boldsymbol l_T=\boldsymbol r_i$ and $\boldsymbol k_T=\boldsymbol q_T+\boldsymbol l_T$, the radiative measurement is $\boldsymbol k_T=\boldsymbol k_n+\boldsymbol k_{\bar n}-\boldsymbol r_o$. 
	Its Fourier identity is
	\begin{align}
		&\delta^{(2)}(\boldsymbol k_T-\boldsymbol k_n-\boldsymbol k_{\bar n} +\boldsymbol r_o)\, \delta^{(2)}(\boldsymbol l_T-\boldsymbol r_i)
		=\int\frac{d^2b_k\,d^2b_l}{(2\pi)^4}\, e^{i\boldsymbol b_k\cdot\boldsymbol k_T} e^{i\boldsymbol b_l\cdot\boldsymbol l_T}
		e^{-i\boldsymbol b_k\cdot (\boldsymbol k_n+\boldsymbol k_{\bar n}-\boldsymbol r_o)} e^{-i\boldsymbol b_l\cdot\boldsymbol r_i}.
		\label{app:scet_measurement}
	\end{align}
	The identity is the origin of the two impact parameters.  
	The phase multiplying each momentum now belongs to a single SCET sector.  
	Its collinear matrix elements are the quark TMD beam functions $B_{n}(x_p,b_k;\mu,\nu)$ and $B_{\bar n}(x_A,b_k;\mu,\nu)$; they contain the active-parton transverse momentum and collinear radiation.  
	The soft matrix element is $S_{\rm soft}(b_k,b_l;\mu,\nu)$, the vacuum expectation value of the two incoming Wilson lines with outside emissions weighted by $\boldsymbol b_k$ and inside emissions by $-\boldsymbol b_l$. 
	Consequently, the impact-parameter integrand is
	\begin{align}
		&e^{i\boldsymbol b_k\cdot\boldsymbol k_T} e^{i\boldsymbol b_l\cdot\boldsymbol l_T} B_{n}(x_p,b_k;\mu,\nu) B_{\bar n}(x_A,b_k;\mu,\nu) S_{\rm soft}(b_k,b_l;\mu,\nu),
		\label{app:scet_factorized_form}
	\end{align}
	multiplied by the hard coefficient $H(Q,\mu)$ and the Born flavor sum. 
	Here $\mu$ and $\nu$ are the virtuality and rapidity renormalization scales. 
	At the natural scale the target beam function is matched onto the CGC sea distribution in Eqs.~\eqref{app:cgc_matching} and \eqref{app:sea_tmd}. 
	Retaining target-fragmentation energy flow replaces that inclusive boundary by $\widetilde{\mathcal B}^{\mathcal C}_{\bar q/A}$ in Eq.~\eqref{eq:target_beam_matching}; the radiative soft function is unchanged. 
	After evolution, its logarithms combine with the hard and beam evolution into $R_{\rm rest}^{\rm pert}$ and $R_{\rm cov}^{\rm pert}$ used in the main text.
	Equations~\eqref{app:scet_measurement} and \eqref{app:scet_factorized_form} give the standard soft and collinear functions.  
	%
	
	In the boson-centered frame, let $p_s^\mu$ be the emitted soft momentum and 	$y_s$ its rapidity.  
	For a symmetric coverage $|y_s|<y_0$, the one-loop soft function refactorizes into the uncovered and covered regions,
	\begin{align}
		S_{\rm soft}^{(1)}(b_k,b_l) ={}&S_{\rm out}^{(1)}(b_k,\mu,\nu) 	+S_{\rm in}^{(1)}(b_l,\mu), \nonumber\\
		S_{\rm out}^{(1)}={}&4g_{s,0}^2C_F\!\int\!\frac{d^dp_s}{(2\pi)^{d-1}}\delta^+(p_s^2)\left(\frac{\nu}{2p_s^0}\right)^\eta		\frac{e^{i\boldsymbol b_k\cdot\boldsymbol p_{sT}}}{p_{sT}^2}\theta(|y_s|-y_0), \nonumber\\
		S_{\rm in}^{(1)}={}&4g_{s,0}^2C_F\!\int\!\frac{d^dp_s}{(2\pi)^{d-1}}\delta^+(p_s^2)\frac{e^{-i\boldsymbol b_l\cdot\boldsymbol p_{sT}}}{p_{sT}^2}\theta(y_0-|y_s|).
		\label{app:scet_soft}
	\end{align}
	Here $d=4-2\epsilon$, $g_{s,0}$ is the bare coupling, and $\delta^+ (p_s^2)=\delta(p_s^2)\theta(p_s^0)$.  
	The regulator $\eta$, with rapidity scale $\nu$, is needed only in the uncovered function.  
	Its pole cancels that of the two beam functions~\cite{Chiu:2012ir}; the covered interval is finite. 
	After renormalization the same integrals give
	\begin{align}
		S_{\rm out}^{(1)}={}&\frac{\alpha_sC_F}{4\pi}\left(2L_k^2-4L_kL_r-\frac{\pi^2}{3}+8y_0L_k\right),	\nonumber\\
		S_{\rm in}^{(1)}={}&-\frac{\alpha_sC_F}{4\pi}\,8y_0L_l ,
		\label{app:scet_soft_result}
	\end{align}
	where $b_0=2e^{-\gamma_E}$, $\mu_{b_k}=b_0/b_k$, and $\mu_{b_l}=b_0/b_l$. The logarithms are $L_k=\ln(\mu^2/\mu_{b_k}^2)$, $L_l=\ln(\mu^2/\mu_{b_l}^2)$, and $L_r=\ln(\nu^2/\mu_{b_k}^2)$.
	The $y_0$ terms cancel at $\boldsymbol b_l=-\boldsymbol b_k$; rotational invariance makes the Sudakov factors depend only on their magnitudes, leaving the inclusive Drell--Yan soft function.  
	Their equal and opposite anomalous dimensions, $\gamma_{{\rm out},0}=-\gamma_{{\rm in},0}=16C_Fy_0$, govern the independent evolution of the two measurement scales.
	
	Beyond one loop, the finite rapidity partition can generate non-global logarithms from correlated soft emissions, which are not included in Eq.~\eqref{app:scet_sudakov}~\cite{Dasgupta:2001sh, Banfi:2002hw, Becher:2015hka, Larkoski:2015zka, Caron-Huot:2015bja, Becher:2016mmh, Becher:2016omr}.
	An all-order analysis of factorization, including Glauber contributions and the structure of associated super-leading logarithms, lies beyond the present accuracy and is left for future work~\cite{Forshaw:2006fk, Gaunt:2014ska, Rothstein:2016bsq, Becher:2021zkk, Banfi:2025mra, Becher:2026kbr, Barcaro:2026dsd}.
	
	The canonical choices are, $\mu_H=Q$ for the hard function, $\mu_B=\mu_{\rm out}=\mu_{b_k}$ for the beam and uncovered soft functions, and $\mu_{\rm in}=\mu_{b_l}$ for the covered soft function.  
	The corresponding rapidity scales are $\nu_B=Q$ and $\nu_{\rm out}=\mu_{b_k}$; the covered soft function requires no rapidity scale. 
	Solving the hard, beam, and soft evolution gives
	\begin{align}
		R_{\rm SCET}^{\rm pert}(b_k,b_l)={}&\frac{C_F}{\pi}
		\Bigg[\int_{\mu_{b_k}^2}^{Q^2}\frac{d\mu^2}{\mu^2}\,\alpha_s(\mu)\left(\ln\frac{Q^2}{\mu^2}-\frac32\right)
		+2y_0\int_{\mu_{b_l}^2}^{\mu_{b_k}^2}\frac{d\mu^2}{\mu^2}\,\alpha_s(\mu)\Bigg].
		\label{app:scet_sudakov}
	\end{align}
	The first integral is the standard Drell--Yan evolution; the second evolves between the two Fourier scales over the measured rapidity length $2y_0$.
	
	In the logarithmic region where the soft interval contains the coverage, the equality with the phase-space derivation is algebraic. 
	Setting $Q_{\rm eff}=Qe^{-y_0}$ and regrouping the two terms in Eq.~\eqref{app:heuristic_split} gives $R_{\rm rest}^{\rm pert}(b_k)+R_{\rm cov}^{\rm pert}(b_l) 	=R_{\rm SCET}^{\rm pert}(b_k,b_l)$, because $\ln(Q^2/Q_{\rm eff}^2)=2y_0$.  
	The Fourier phase in the SCET factorization is $e^{i\boldsymbol b_k\cdot\boldsymbol k_T} e^{i\boldsymbol b_l\cdot\boldsymbol l_T}$. 
	At fixed $\boldsymbol q_T$, $\boldsymbol k_T=\boldsymbol q_T+\boldsymbol l_T$, so the finite $l_T$ cut produces exactly the Bessel kernel in Eq.~\eqref{app:veto_kernel}.  
	For a general laboratory interval, the rapidity step functions in Eq.~\eqref{app:scet_soft} are replaced by the corresponding coverage function.  
	After the beam--soft rapidity cancellation their overlap with $|y_s|<L_\mu$ is $\Delta\eta_{\rm cov}(\mu;y_Z^{\rm lab})$; this includes the detector-edge turn-on and recovers Eqs.~\eqref{eq:covered_rapidity}--\eqref{eq:residual_radiator}.  
	The SCET and rapidity-phase-space derivations are therefore the operator and emission-level forms of the same Sudakov evolution.

\end{document}